\documentclass[a4paper,fleqn,usenatbib]{mnras}

\usepackage{newtxtext,newtxmath}
\usepackage[T1]{fontenc}
\usepackage{ae,aecompl}

\usepackage{graphicx}	% Including figure files
\usepackage{amsmath}	% Advanced maths commands
\usepackage{amssymb}	% Extra maths symbols

\newcommand{\HI}{\mathrm{HI}}
\newcommand{\HII}{\mathrm{HII}}
\newcommand{\HeI}{\mathrm{HeI}}
\newcommand{\HeII}{\mathrm{HeII}}
\newcommand{\HeIII}{\mathrm{HeIII}}
\newcommand{\GHI}{\Gamma_{\rm HI}}

\newcommand{\lya}{Ly$\alpha$ }

\newcommand{\teff}{\tau_\mathrm{eff}}

\defcitealias{Eilers17}{E17}

\title[Non-equilibrium quasar proximity zones]{Time-dependent behaviour of quasar proximity zones at $z\sim6$}

\author[F. B. Davies et al.]{
Frederick B. Davies,$^{1}$\thanks{E-mail: davies@physics.ucsb.edu (FBD)}
Joseph F. Hennawi,$^{1}$
Anna-Christina Eilers$^{2}$\thanks{IMPRS Fellow}
\\
$^{1}$Department of Physics, University of California, Santa Barbara, CA 93106-9530, USA\\
$^{2}$Max Planck Institute for Astronomy, K\"onigstuhl 17, 69117 Heidelberg, Germany\\
}

\date{Accepted XXX. Received YYY; in original form ZZZ}

\pubyear{2019}

\begin{document}
\label{firstpage}
\pagerange{\pageref{firstpage}--\pageref{lastpage}}
\maketitle

\begin{abstract}
  Since the discovery of $z\sim 6$ quasars two decades ago, studies of
  their Ly$\alpha$-transparent proximity zones have largely focused on
  their utility as a probe of cosmic reionization. But even when
  in a highly ionized intergalactic medium, these zones provide a rich laboratory for determining
  the  timescales that govern quasar activity and the concomitant growth of
  their supermassive black holes.  In this work, we use a suite of
  1D radiative transfer simulations of quasar proximity zones to explore
  their time-dependent behaviour for activity timescales from $\sim10^3$--$10^8$ years. 
  The sizes of the simulated proximity zones, as quantified by
  the distance at which the smoothed Ly$\alpha$ transmission drops
  below 10\% (denoted $R_p$), are in excellent agreement with
  observations, with the exception of a handful of particularly small
  zones that have been attributed to extremely short $\lesssim 10^4$
  lifetimes. We develop a physically motivated semi-analytic
  model of proximity zones which captures the bulk of their
  equilibrium and non-equilibrium behaviour, and use this model to
  investigate how quasar variability on $\la10^5$ year timescales
  is imprinted on the distribution of observed proximity zone
  sizes. We show that large variations in the ionizing luminosity of
  quasars on timescales of $\la10^4$ years are disfavored based on the
  good agreement between the observed distribution of $R_p$ and our
  model prediction based on ``lightbulb'' (i.e. steady constant
  emission) light curves.
\end{abstract}


%%%%%%%%%%%%%%%%%%%%%%%%%%%%%%%%%%%%%%%%%%%%%%%%%%

%%%%%%%%%%%%%%%%% BODY OF PAPER %%%%%%%%%%%%%%%%%%

\section{Introduction}

The ``proximity effect" is the well-known tendency for decreased \lya forest absorption in the intergalactic medium (IGM) close to luminous quasars along the line of sight (e.g. \citealt{Bajtlik88}). This decrease in absorption is due to the strong ionizing radiation field produced by the quasar itself,
which can greatly outshine the intergalactic ionizing background at small physical separations.

A related effect was predicted for quasars observed during the epoch of reionization (e.g. \citealt{SG87,MR00,CH00}, and see \citealt{Hogan97} for a He\,{\small II} analogy),
wherein the cumulative ionizing photon output from a luminous quasar carves out a transparent H\,{\small II} region along the line of sight. The size of this ionized region, in the absence of recombinations, is given by
\begin{equation} \label{eqn:rion}
R_{\rm ion} = \left( \frac{3\dot{N}_{\rm ion} t_{\rm q}}{4\pi n_{\rm H} x_{\rm HI}} \right)^{1/3},
\end{equation}
where $\dot{N}_{\rm ion}$ is the emission rate of ionizing photons, $t_{\rm q}$ is the age of the quasar, $n_{\rm H}$ is the (average) number density of hydrogen atoms, and $x_{\rm HI}$ is the neutral fraction of the IGM. If luminous quasars shine for $>10^6$ years, $R_{\rm ion}$ can reach scales of several proper Mpc, corresponding to a few thousand km s$^{-1}$ along the line of sight in the $z\gtrsim6$ \lya forest. 
 
Observations of transparent proximity zones around quasars at $z\ga6$
terminated by opaque Gunn-Peterson troughs \citep{Becker01,White03}
appeared to confirm the picture of quasar-ionized bubbles in a neutral
IGM, and led to claims of constraints on the reionization history of
the Universe \citep{WL04} and measurement of the evolving residual IGM neutral
hydrogen fraction after reionization was complete
\citep{Fan06,Carilli10}.
The properties of the first large sample of
$z\ga6$ quasar spectra compiled by \citet{Fan06} led them to define
the size of the proximity zone, denoted by $R_p$, as the first
location where the \lya forest spectrum drops below 10\% transmitted
flux when smoothed to 20\,{\AA} in the observed frame, roughly
corresponding to a physical scale of 1 proper Mpc at
$z\sim6$. 

These transparent proximity zones observed at $z\ga 6$ can be
interpreted in the context of the ionized bubbles
expanding around quasars during the epoch of reionization, as described above.
In this model, the IGM interior to $R_{\rm ion}$ is
assumed to be transparent while the external IGM is fully opaque, thus $R_p$ is identified with the
ionization front radius $R_{\rm ion}$. The scaling of $R_p$
with various parameters can then be read directly from
equation~(\ref{eqn:rion}): $R_p\propto t_{\rm q}^{1/3}$, $R_p\propto
\dot{N}_{\rm ion}^{1/3}$, and $R_p\propto x_{\rm HI}^{-1/3}$. The
$\dot{N}_{\rm ion}$ scaling in particular has been used to ``correct"
measurements of $R_p$ onto a fixed quasar luminosity scale
\citep{Fan06,Carilli10,Venemans15}.

However, as noted by \cite{BH07}, the \lya forest at $z\sim6$ becomes 
opaque at neutral fractions as low as $x_{\rm HI}\sim10^{-4}$, and thus the \emph{observed} proximity zone (as defined by $R_p$) may be truncated well before any actual ionization front (i.e. at distances less than $R_{\rm ion}$). Indeed, \citet{BH07} found that if the IGM is already highly ionized, consistent with measurements of the mean transmitted flux in the IGM at $z\sim6$, the proximity zone sizes will be roughly $R_p\sim5$--$10$ proper Mpc, in agreement with the existing observations which had been previously interpreted in terms of the location of $R_{\rm ion}$ in a neutral IGM. In this highly ionized regime, $R_p$ is no longer directly connected to reionization. Instead, $R_p$ reflects the distance at which the enhanced ionizing flux from the quasar (combined with the ionizing background radiation) becomes weak enough for Ly$\alpha$ transmission to fall below 10\%.

In the \citet{BH07} model, the \lya transmission of the IGM is
approximated by the Gunn-Peterson (GP) optical depth (e.g. \citealt{Weinberg97}),
\begin{equation}
\tau_{\rm GP} = \frac{\sigma_\alpha c n_{\rm H} x_{\rm HI}}{H(z)},
\end{equation}
where $\sigma_\alpha$ is the Ly$\alpha$ scattering cross-section, $c$ is the speed of light, and $H(z)$ is the Hubble parameter. Allowing $x_{\rm HI}$ to vary along the line of sight due to the enhanced photoionization rate of the quasar, the \emph{observed} proximity zone size, assuming a transmission threshold of 10\%, is given by
\begin{eqnarray} \label{eqn:rpbh07}
R_p = \frac{3.14\ \rm proper\ Mpc}{\Delta_{\rm lim}} && \left(\frac{\dot{N}_{\rm ion}}{2\times10^{57}\ {\rm s}^{-1}}\right)^{1/2} \\ \nonumber &&\times \left(\frac{T}{2\times10^4\ {\rm K}}\right)^{0.35} \left(\frac{1+z}{7}\right)^{-9/4},
\end{eqnarray}
where $T$ is the temperature of the IGM and $\Delta_{\rm lim}$ is an ``effective" IGM overdensity which represents an unknown renormalization from the GP optical depth to the true \emph{effective} optical depth of the Ly$\alpha$ forest implicit in the expression. In this case, $R_p$ scales as $\dot{N}_{\rm ion}^{1/2}$, but is otherwise independent of the residual IGM neutral fraction as the quasar is assumed to dominate over the ionizing background.
Note that the IGM is assumed to be in ionization equilibrium, so $R_p$ is also independent of the quasar lifetime.
In this model, however, it is difficult to explain the rapid evolution in $R_p$ measured by \citet{Fan06} and \citet{Carilli10} (see also \citealt{Venemans15}).

For quasars at $z\sim6$, it is quite likely that the IGM around quasar hosts is ionized before the quasar turns on. The neutral fraction of the Universe at $z\sim6$ is $\la0.2$ \citep{McGreer15} and the standard inside-out model of reionization \citep{Furlanetto04} predicts that the large-scale environment of massive dark matter halos was ionized relatively early \citep{AA07,Lidz07}. It is worth noting, however, that the ionized IGM regime for $R_p$ will still apply to a partially
neutral IGM if the pre-existing ionized region is significantly larger than what $R_p$ would have been in the absence of neutral gas \citep{BH07,Maselli09}. In the following we will assume that, for the purposes of analyzing quasar proximity zones via $R_p$, the Universe is fully reionized.

Recently, \citet[][henceforth \citetalias{Eilers17}]{Eilers17} compiled a large number
of high-redshift quasar spectra from the Keck Observatory Archive and
new Keck observations to perform a homogeneous measurement of $R_p$
with higher signal-to-noise spectra, consistently defined $M_{1450}$,
and principal component analysis continuum estimation. From their
sample of 31 quasars covering $5.77 \leq z \leq 6.54$, they measured a
very shallow evolution of $R_p$ with redshift, inconsistent with
previous works but consistent with radiative transfer simulations of
quasars in the post-reionization IGM. They also discovered a handful
of quasars with much smaller proximity zones than predicted, and
suggested that these small zones could be due
to non-equilibrium ionization of the IGM close to the quasar, requiring the duration of current quasar activity to be
shorter than $\sim10^{4-5}$ years (see also \citealt{Eilers18J1335}). 
Additional evidence for non-equilibrium effects in quasar proximity zones
 has also recently been measured in the He\,{\small II} Ly$\alpha$ forest \citep{Khrykin19}.
However, this short timescale behaviour was only studied in the context of a standard ``lightbulb" model of quasar activity,
wherein the quasar switched on at some time in the past and has maintained the same luminosity since then.

In this work, we expand upon previous explorations of $z\ga6$ quasar
proximity zones from a theoretical perspective, in light of the
observations in \citetalias{Eilers17}. In \S~2, we describe our 1D radiative transfer
simulations and summarize the physical and observational picture of
quasar proximity zones as modeled by skewers through a large-volume
high-resolution hydrodynamical simulation. In \S~3, using the behaviour
of the \lya forest in our hydrodynamical simulation as a guide, we
build a new semi-analytic model for the \lya transmission profile in
quasar proximity zones in a highly ionized IGM.
In \S~4 we explore how quasar variability on timescales less than $10^5$ years can be imprinted on the proximity zone profile. Finally, we conclude in \S~5.

We assume a $\Lambda$CDM cosmology with $\Omega_m=0.3$,
$\Omega_\Lambda=0.7$, $\Omega_{\rm b}=0.047$, and $h = 0.685$, consistent
with the CMB constraints from \citet{Planck15}.

\begin{figure*}
\begin{center}
\resizebox{14cm}{!}{\includegraphics[trim={1.0em 1em 1.0em 1em},clip]{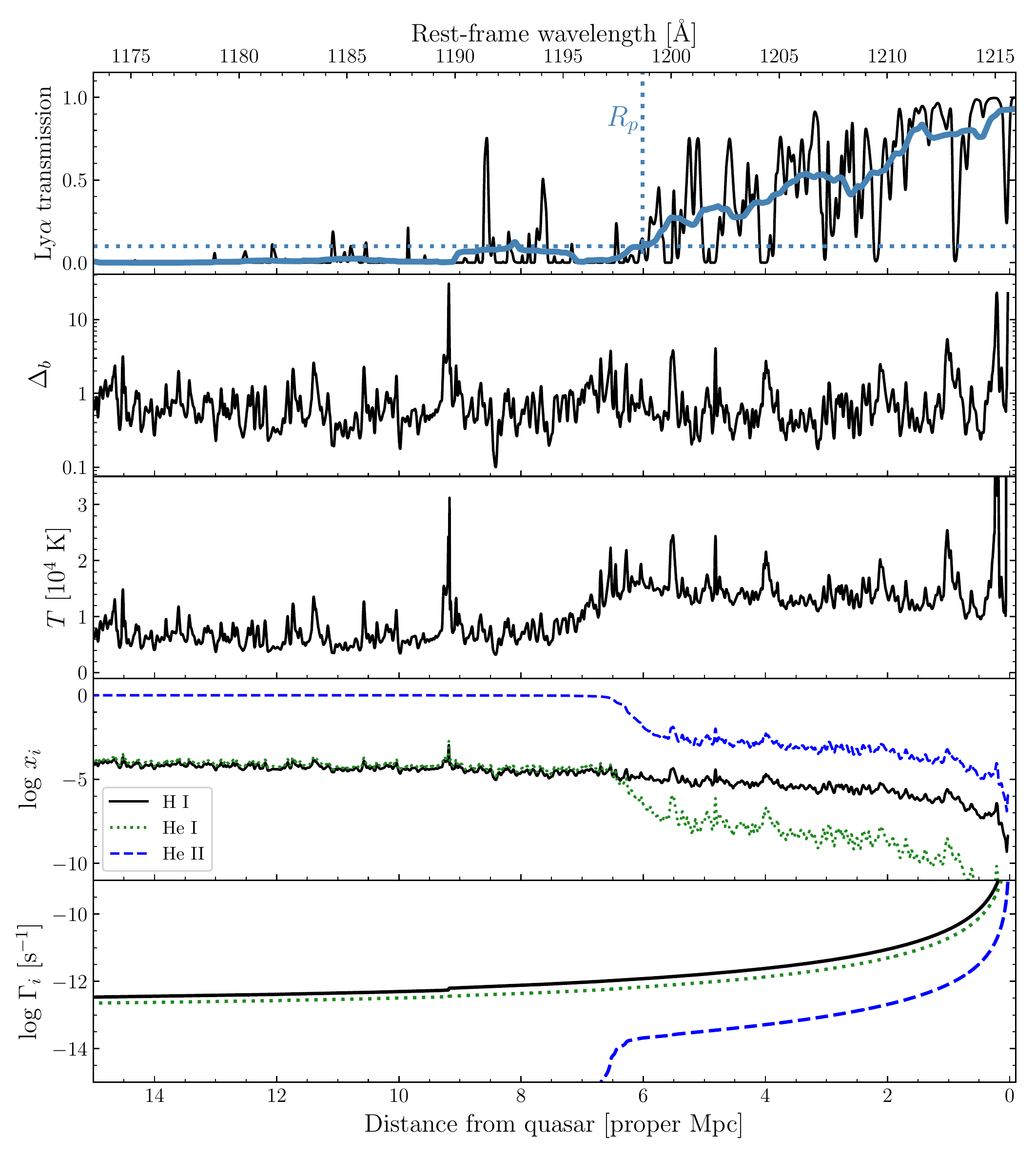}}
\end{center}
\caption{Example of a 1D radiative transfer simulation used in this work. The black curve in the top panel shows the Ly$\alpha$ transmission spectrum, with a transmitted flux of 10\% marked by the horizontal dotted line. The blue curve shows the spectrum smoothed by 20\,{\AA} in the observed frame, and the vertical dotted line indicates when the smoothed spectrum first drops below 10\%, defined to be the location of $R_p$. The remaining panels from top to bottom show the baryon overdensity, the gas temperature, the fractions of H\,{\small I}/\ion{He}{1}/He\,{\small II}, and the photoionization rates of H\,{\small I}/\ion{He}{1}/He\,{\small II}. }
\label{fig:rt_ex}
\end{figure*}

\section{Radiative Transfer modelling of Proximity Zones}

Our radiative transfer modelling of quasar proximity zones closely follows the methods described in \citet{Davies16} and \citetalias{Eilers17}, which we summarize briefly below. 

\subsection{Hydrodynamical simulation}\label{sec:hydro}

We use hydrodynamical simulations skewers drawn from a Nyx Eulerian grid hydrodynamical simulation \citep{Almgren13,Lukic15} 100 Mpc$/h$ on a side with $4096^3$ dark matter particles and baryon grid elements. We draw $900$ skewers from the $z=5.5$, $6.0$, and $6.5$ outputs of the simulation starting from the centers of the $150$ most massive dark matter halos at each redshift ($M_h\ga10^{11.5}$ M$_\odot$). While the Eulerian grid does not resolve the very dense regions ($\la r_{\rm vir}$) associated with the small-scale environment ($R<R_{\rm vir}$) of these massive halos particularly well, the lower-density IGM that gives rise to transmission in the \lya forest on large scales is well-converged (J. O\~{n}orbe, priv. comm.), and the rarity of proximate neutral absorbers in lower redshift quasars (e.g. \citealt{Prochaska08}) suggests that the material on such scales does not usually affect the proximity zone (although notable exceptions exist, see \citealt{D'Odorico18,Banados19}).%,Davies19c}).

The reionization and thermal history of the hydrodynamical simulation was set by the \citep{HM12} model for the evolution of the ionizing background, leading to reionization at $z>10$ with an associated heat input of $\sim10^4$ K \citep{Lukic15,Onorbe17}. 
The resulting thermal state of the IGM at $z=6$ is well-described by a power law relationship between temperature ($T$) and overdensity ($\Delta\equiv\rho/\bar{\rho}$) with $T=T_0\Delta^{\gamma-1}$, $T_0\sim7500$ K, and $\gamma\sim1.5$. The thermal state of the ambient IGM sets the normalization of the relationship between the ionization rate and the mean transmission through the \lya forest, which we discuss later in \S~\ref{sec:eq}.

\subsection{1D radiative transfer}\label{sec:rt}

To compute the effect of quasar ionizing radiation on the IGM, we perform 1D radiative transfer (RT) along the simulations skewers described above. We use the 1D RT code developed in \citet{Davies16}, originally based on the method described in \citet{BH07}, with some minor adjustments as described in \citetalias{Eilers17}. We summarize the basic model below, but the code is described in much more detail in \citet{Davies16}.

The code solves the following time-dependent equations for ionized H and He species,
\begin{eqnarray}
\frac{dn_\HII}{dt} &=& n_\HI (\GHI+n_e\Gamma^e_\HI) - n_e n_\HII \alpha^{\rm A}_\HII \label{eqn:nhi}\\
\frac{dn_\HeII}{dt} &=& n_\HeI (\Gamma_\HeI+n_e\Gamma^e_\HeI) + n_\HeIII n_e \alpha^{\rm A}_\HeIII\\
&& - n_\HeII (\Gamma_\HeII + n_e \Gamma^e_\HeII + n_e \alpha^{\rm A}_\HeII) \\
\frac{dn_\HeIII}{dt} &=& n_\HeII (\Gamma_\HeII + n_e \Gamma^e_\HeII) - n_e n_\HeIII \alpha^{\rm A}_\HeIII
\end{eqnarray}
where $n_i$ are the number densities of species $i$, $\Gamma_i$ are
the photoionization rates of species $i$, $\Gamma^e_i$ are the collisional ionization rates from \citet{Theuns98}, and $\alpha^{\rm A}_i$ are the
Case A recombination rates of species $i$ from \citet{HG97}. The photoionization rates include secondary ionizations by energetic photoelectrons
\citep{SvS85}, for which we use the results of \citet{FJS10}.
The quasar SED is assumed to follow the
\citet{Lusso15} radio-quiet SED, allowing one
to convert from $M_{1450}$ to the specific luminosity $L_\nu$
at the hydrogen ionizing edge, beyond which we assume a power law
spectrum with $L_\nu \propto \nu^{-1.7}$. The neutral species and the
electron density are recovered by the closing
conditions
\begin{eqnarray}
n_\HI &=& n_{\rm H} - n_\HII \\
n_\HeI &=& n_{\rm He} - n_\HeII - n_\HeIII \\
n_e &=& n_\HII + n_\HeII + 2n_\HeIII.
\end{eqnarray}
 Finally, the gas temperature $T$ is computed by solving
\begin{equation}
\frac{dT}{dt} = \frac{(\gamma_{\rm ad}-1)\mu m_{\rm H}}{k_{\rm B}\rho}(\mathcal{H}-\Lambda)-2H(z)T-\frac{T}{n}\frac{dn_e}{dt},
\end{equation}
where $\gamma_{\rm ad}=5/3$ is the adiabatic index,
$\mu$ is the mean molecular weight, $\rho$ is the gas density, $\mathcal{H}$
is the heating rate, $\Lambda$ is the cooling rate, and $n$ is the total number density of all species. The first term represents the balance between the various gas-phase heating and cooling rates, the second term represents the adiabatic cooling due to the expansion of the Universe, and the last term evenly distributes thermal energy between species as the number of particles changes.

For simplicity we assume that $z$ is constant with time in every gas cell during the radiative transfer calculation, however in an attempt to produce slightly more accurate transmission profiles on large scales, we allow the redshift of each cell -- and the physical gas densities, assuming $\rho\propto(1+z)^{3}$ -- to vary as a function of distance from the quasar according to the cosmological line increment $dl/dz=c/[(1+z)H(z)]$. In practice this has a very modest effect on the mock spectra.

In Figure~\ref{fig:rt_ex}, we show the physical properties and Ly$\alpha$ transmission spectrum of a simulated skewer for a $z=6$ quasar with $M_{1450}=-27$ which has been on for $t_{\rm q}=10^{7.5}$ years, the fiducial quasar age assumed in \citetalias{Eilers17}. The excess Ly$\alpha$ transmission due to the proximity effect is clearly evident out to distances of nearly 10 proper Mpc, with excess heat from He\,{\small II} ionization visible out to $\sim6$ proper Mpc.

\subsection{The proximity zone size, $R_p$}\label{sec:rtrp}

We measure $R_p$ in the RT simulations analogously to the observations. The Ly$\alpha$ transmission spectra are smoothed by a 20 \AA\ top hat filter, and then the first location where the transmission falls below 10\% (starting from $R=0$) is recorded as $R_p$. The blue curve in the top panel of Figure~\ref{fig:rt_ex} shows the smoothed spectrum and the location of $R_p$. For this particular simulated skewer, $R_p\approx6.0$ proper Mpc, which is typical for quasars of this luminosity at $z\sim6$ \citepalias{Eilers17}.

\begin{figure}
\begin{center}
\resizebox{8.5cm}{!}{\includegraphics[trim={1em 0.5em 1em 1em},clip]{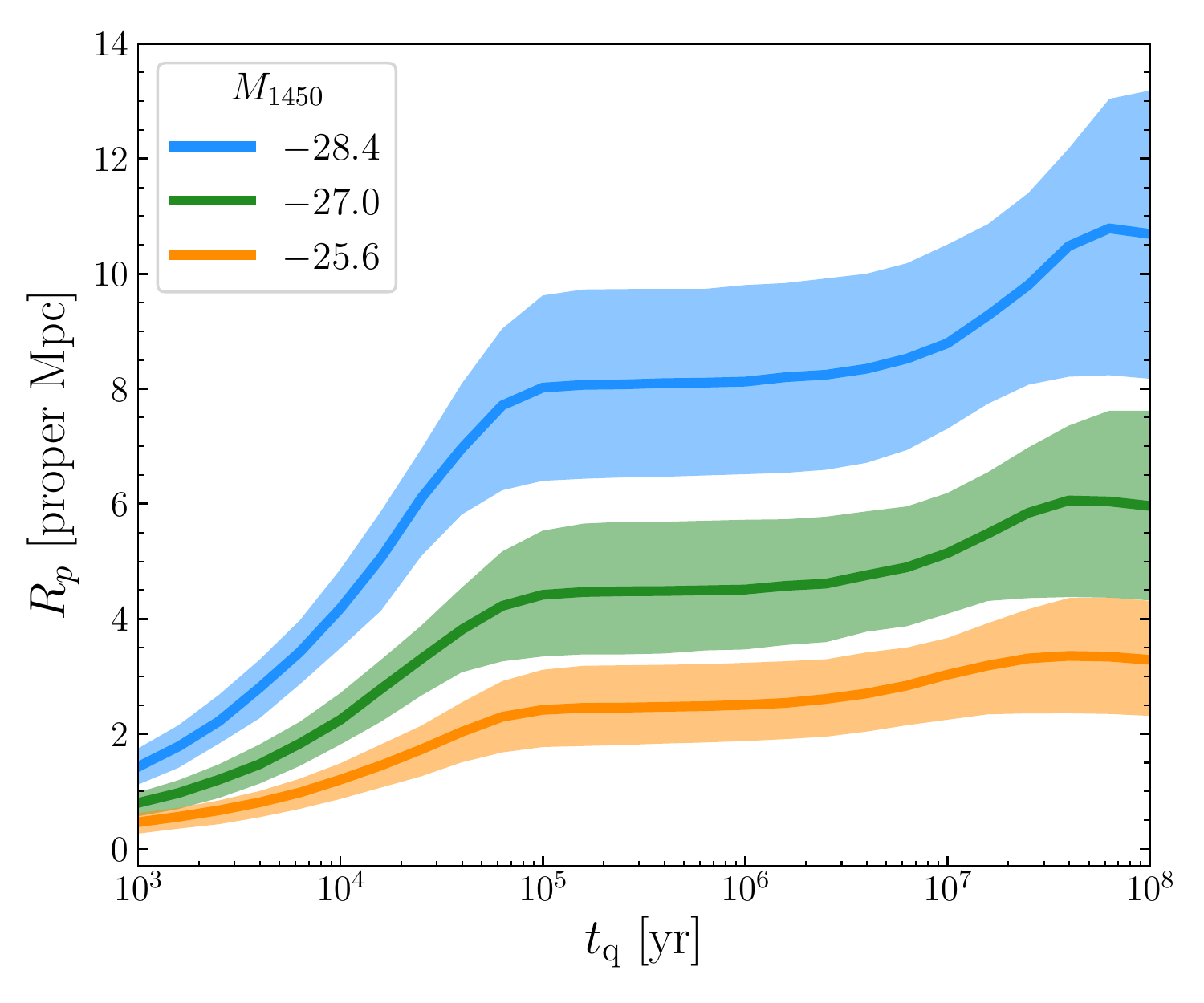}}
\end{center}
\caption{Evolution of $R_p$ in the RT simulations for $M_{1450}=-25.6$ (orange), $-27.0$ (green), and $-28.4$ (blue). The solid curves show the median $R_p$, while the shaded regions show the 16--84th percentile range at each $t_{\rm q}$.}
\label{fig:rt_rp_tq}
\end{figure}

\begin{figure}
\begin{center}
\resizebox{8.5cm}{!}{\includegraphics[trim={1em 1em 1em 1em},clip]{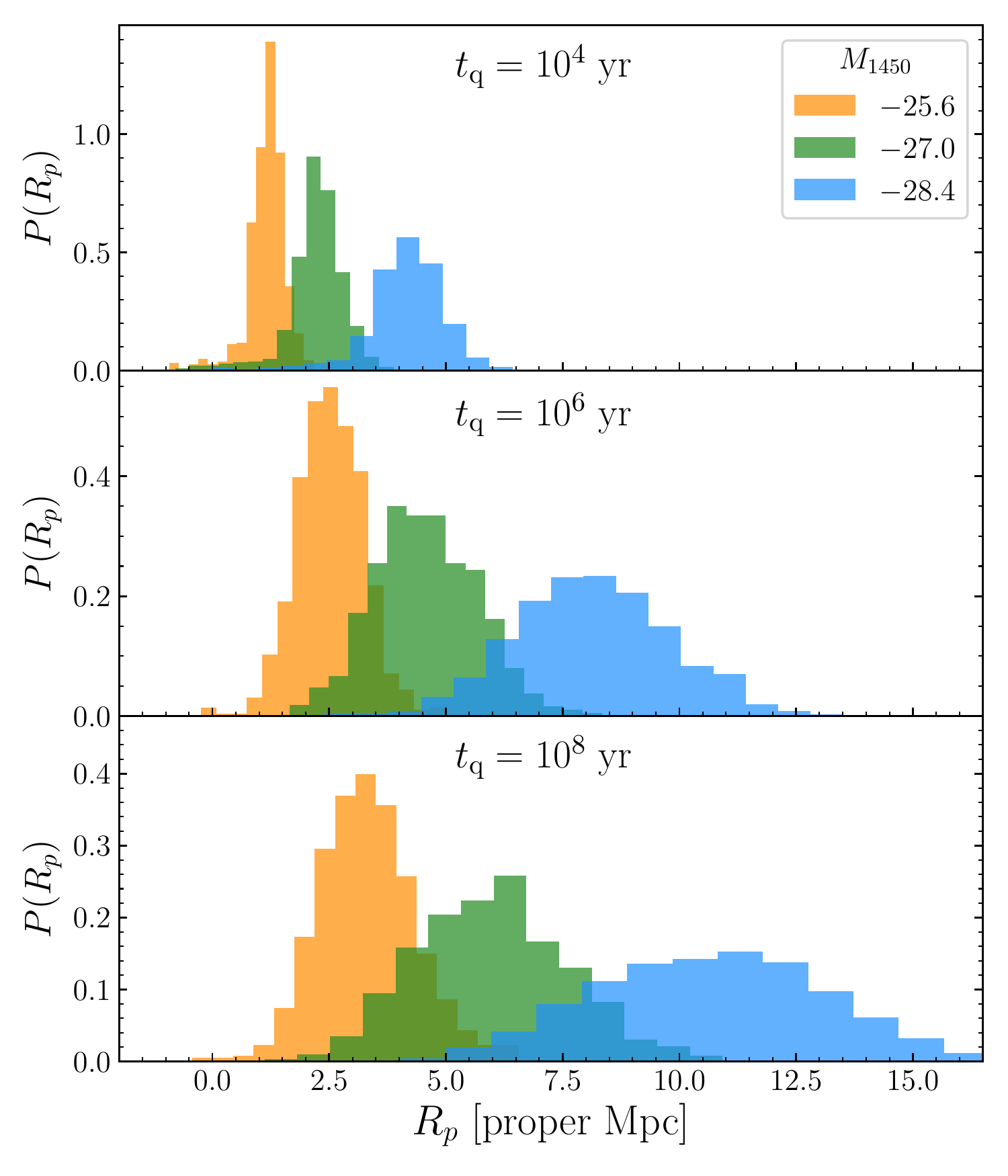}}
\end{center}
\caption{Distribution of $R_p$ in the RT simulations for different combinations of $t_{\rm q}=10^4$, $10^6$, and $10^8$ yr, from top to bottom, and $M_{1450}=-25.6$ (orange), $-27.0$ (green), and $-28.4$ (blue).}
\label{fig:rt_rp_hist}
\end{figure}

\begin{figure}
\begin{center}
\resizebox{8.5cm}{!}{\includegraphics[trim={1em 0.5em 1em 1em},clip]{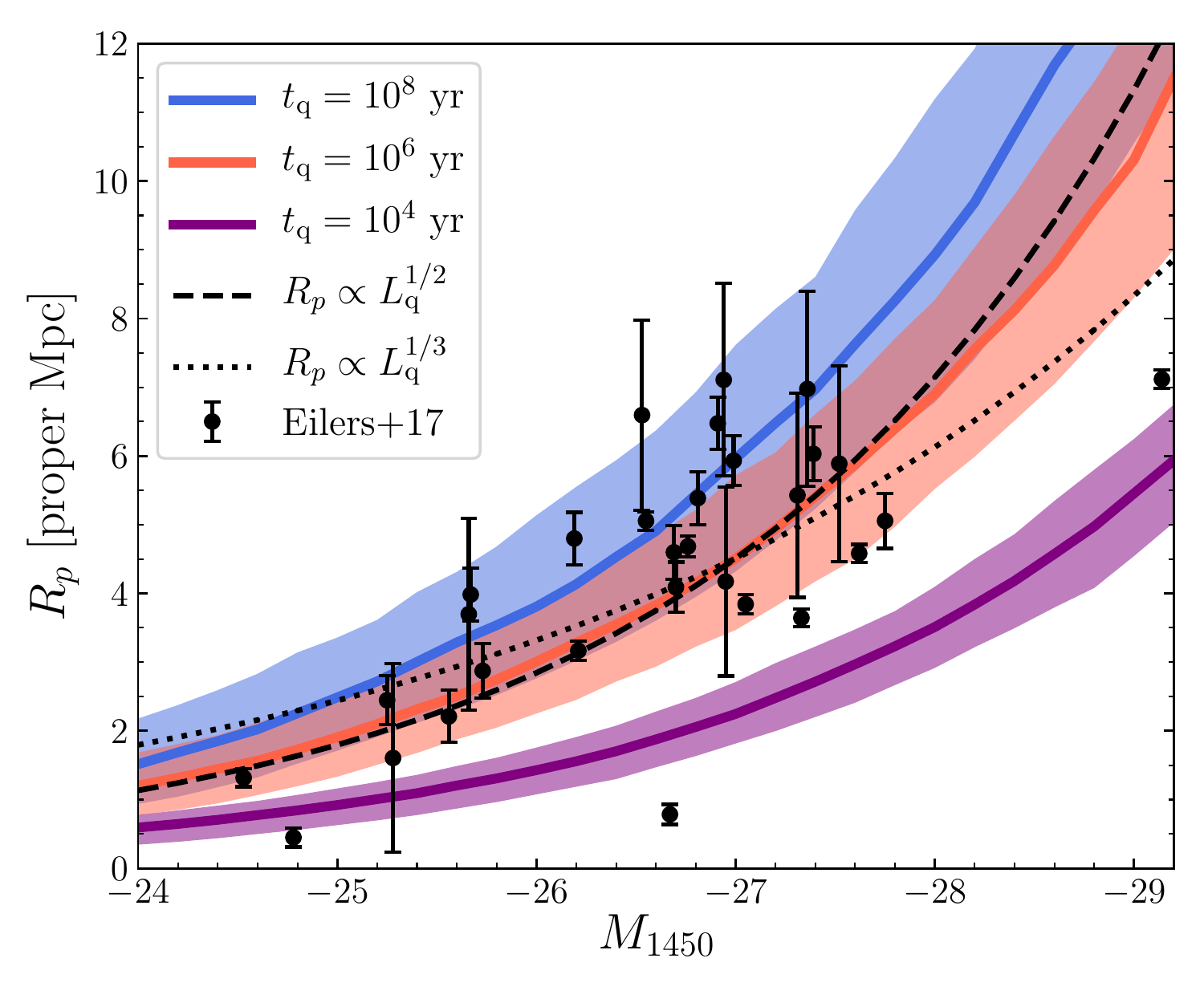}}
\end{center}
\caption{Median $R_p$ (solid curves) and its 16--84th percentile scatter (shaded regions) in the RT simulations as a function of $M_{1450}$ for $t_{\rm q}=10^4$ (purple), $10^6$ (red), and $10^8$ years (blue). Measured $R_p$ from \citetalias{Eilers17} are shown as black points with quasar systemic redshift uncertainties.}
\label{fig:rt_rp_mag}
\end{figure}

\begin{figure}
\begin{center}
\resizebox{8.5cm}{!}{\includegraphics[trim={0.5em 1em 3.6em 0em},clip]{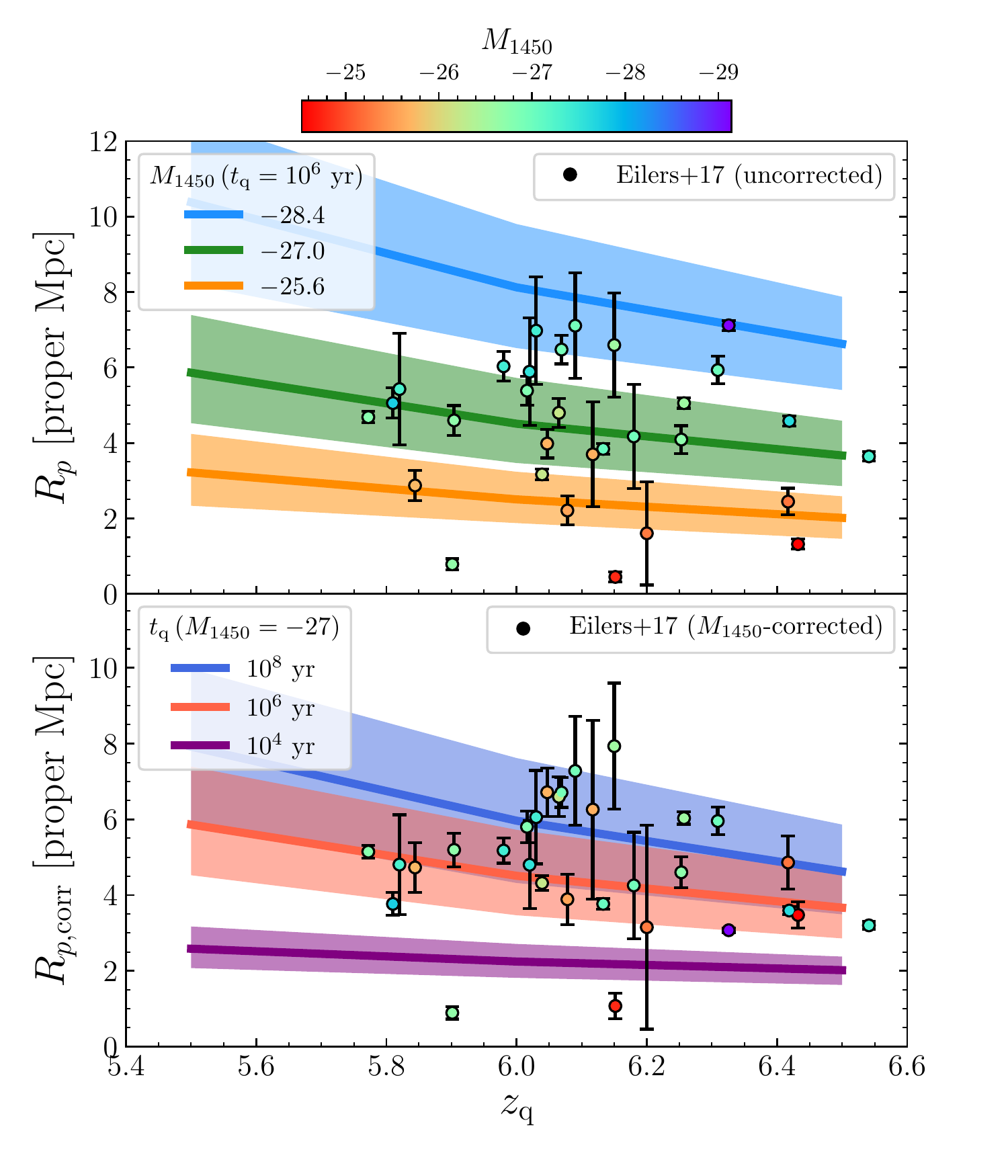}}
\end{center}
\caption{Evolution of $R_p$ with redshift. Top: Median $R_p$ (solid curves) and its 16--84th percentile scatter (shaded regions) in the RT simulations as a function of $z_{\rm q}$ for $M_{1450}=-25.6$ (orange), $-27.0$ (green), and $-28.4$ (blue) at a fixed quasar age of $t_{\rm q}=10^6$ yr. The RT simulations were only run at $z=5.5,6.0,6.5$; the curves and shaded regions represent a linear interpolation between the distributions at these three redshifts. Measured $R_p$ from \citetalias{Eilers17} are shown as points, coloured by their corresponding $M_{1450}$, with error bars showing quasar systemic redshift uncertainties. Bottom: Similar to the top panel but now showing $R_p$ in the RT simulations for $t_{\rm q}=10^4$ yr (purple), $10^6$ yr (red), and $10^8$ yr (blue) for a quasar with $M_{1450}=-27.0$. $M_{1450}$-corrected $R_p$ from \citetalias{Eilers17}, $R_{p,{\rm corr}}$, are shown as black points with (similarly corrected) systemic redshift uncertainties.}
\label{fig:rt_rp_redshift}
\end{figure}

In Figure~\ref{fig:rt_rp_tq} we show how $R_p$ evolves as a function of time for three different quasar luminosities. Two distinct regimes of $R_p$ growth are evident: an initial rapid rise for $t_{\rm q} \lesssim 10^5$ yr, and a modest increase at $10^7 \lesssim t_{\rm q} \lesssim 10^8$ yr. As discussed in \citetalias{Eilers17} and \citet{Eilers18J1335}, the first of these increases is due to the finite response time of hydrogen
to the ionizing photons from the quasar. As described in \citet{McQuinn09b}, and discussed more in \S~\ref{sec:sam}, the IGM responds to changes in ionizing radiation over a characteristic timescale of $t_{\rm eq}\sim \Gamma^{-1}$, which for hydrogen in $z\sim6$ quasar proximity zones corresponds to $\sim3\times10^4$ years \citepalias{Eilers17}. The second increase is due to He\,{\small II} photoionization heating, which we discuss further below.

The shaded regions in Figure~\ref{fig:rt_rp_tq} show the central 68\% scatter of $R_p$ between different IGM skewers at fixed $t_{\rm q}$. This scatter arises from
 fluctuations in the density field causing the (smoothed) transmitted flux to drop below 10\% at different locations in each spectrum. We show the full distributions of $R_p$ for three representative quasar ages in Figure~\ref{fig:rt_rp_hist}, demonstrating that as $R_p$ increases, the scatter in $R_p$ increases as well. The distributions of $R_p$ are largely Gaussian in shape, with weak tails to lower and higher values at early and late times, respectively.

Figure~\ref{fig:rt_rp_mag} shows the dependence of $R_p$ on quasar luminosity in our simulations compared to the measurements from
\citetalias{Eilers17}. The majority of quasar spectra analyzed by \citetalias{Eilers17} have $R_p$ consistent with $t_{\rm q}\ga10^6$ yr, with a few notable exceptions that may indicate much younger ages (see also \citealt{Eilers18J1335}). Common theoretical luminosity scalings of $R_p$ are shown by the dashed ($R_p\propto \dot{N}_{\rm ion}^{1/2}$) and dotted curves ($R_p\propto \dot{N}_{\rm ion}^{1/3}$), anchored to the $t_{\rm q}=10^6$ yr simulations at $M_{1450}=-27$. The $\dot{N}_{\rm ion}^{1/2}$ scaling arises from the assumption that $R_p$ occurs at a fixed quasar ionizing flux \citep{BH07}, while the $\dot{N}_{\rm ion}^{1/3}$ scaling comes from balancing the number of emitted ionizing photons with the number of hydrogen atoms in the IGM (equation~\ref{eqn:rion}, \citealt{CH00}). The scaling in our RT simulations of a highly ionized IGM roughly agree with the $\dot{N}_{\rm ion}^{1/2}$, but subtly disagree due to the effect of IGM density fluctuations and He\,{\small II} reionization heating, which we discuss further below.

Figure~\ref{fig:rt_rp_redshift} shows how $R_p$ evolves with redshift
with a fixed UV background intensity. The redshift evolution of $R_p$ is then driven by cosmological density evolution due to the expansion of the Universe as well as structure formation.
We improve upon the redshift
evolution of the simulations presented in \citetalias{Eilers17} by showing RT simulations using
hydrodynamical simulation outputs at $z=5.5$ and $z=6.5$ rather than
simply re-scaling the $z=6.0$ density skewers by $(1+z)^3$. The
evolution of $R_p$ in the simulations is found to roughly follow
$R_p\propto(1+z)^{-3.2}$,
which is fairly shallow over the redshift
range covered by \citetalias{Eilers17}, consistent with the non-evolution implied by
their measurements. In the top panel of
Figure~\ref{fig:rt_rp_redshift} we show the ``raw"
$R_p$ measurements
from \citetalias{Eilers17} compared to our simulations as a function of $M_{1450}$, with
a fixed $t_{\rm q}=10^6$ yr. In the bottom panel, we instead show
varying $t_{\rm q}$ at fixed $M_{1450}=-27.0$ in the simulations,
compared to the ``corrected" $R_p$ values from \citetalias{Eilers17}, $R_{p,{\rm
    corr}}$, which have been rescaled 
so as to remove the dependence
on quasar luminosity. While \citet{Fan06} and \citet{Carilli10} made
these corrections assuming $R_p\propto \dot{N}_{\rm ion}^{1/3}$, \citetalias{Eilers17} instead
used  $R_p\propto \dot{N}_{\rm ion}^{1/2.35}$, a relationship derived from our RT
simulations. However, it is worth noting that
this relationship was derived for $t_{\rm q}=10^{7.5}$ yr,
but we find that different quasar ages result in subtly different
dependences. For example, for $t_{\rm q}=10^6$ yr we
find $R_p \propto \dot{N}_{\rm ion}^{1/2.2}$ is a better fit to the behaviour
of the median $R_p$.

In general, the measured properties of quasar proximity zones -- as traced by $R_p$ -- 
appear to be in good agreement with our RT simulations. This agreement is not trivial, as
it does not represent any kind of arbitrary tuning of, e.g., the IGM density field or
of the quasar SEDs to produce more or fewer ionizing photons. 

\subsection{Impact of He\,{\small II} reionization heating} \label{sec:helium}

Prior to the quasar turning on, the helium in the simulation is almost entirely in the He\,{\small II} ionization state. The reionization of He\,{\small II} by the quasar heats the gas by $\sim10^4$ K within a radius given by the He\,{\small II}
analogy of equation~(\ref{eqn:rion}),
\begin{equation} \label{eqn:rion_he}
R_{\rm ion}^{\rm He} = \left( \frac{3\dot{N}_{\rm ion}^{\rm He\, II} t_{\rm q}}{4\pi n_{\rm He} x_{\rm He\,II}} \right)^{1/3},
\end{equation}
where $\dot{N}_{\rm ion}^{\rm He\, II}=4^{-1.7} \dot{N}_{\rm ion}$ is the number of He\,{\small II}-ionizing photons emitted by the quasar for our choice of quasar SED, $n_{\rm He}$ is the helium number density, and $x_{\rm He\,II}$ is the He\,{\small II} fraction (which should be close to unity at $z>5$). This excess heat, known as the thermal proximity effect \citep{Meiksin10,Bolton10,Bolton12,Khrykin17},
reduces the H\,{\small I} fraction of the IGM somewhat ($x_{\rm HI}\propto \Gamma_{\rm HI}^{-1} T^{-0.7}$), and thus leads to extra transmission at $R<R_{\rm ion}^{\rm He}$. The increase in Ly$\alpha$ transmission, and corresponding increase in $R_p$
at late times
is then due to $R_{\rm ion}^{\rm He}$ reaching the unheated $R_p$. 

\begin{figure}
\begin{center}
\resizebox{8.5cm}{!}{\includegraphics[trim={1em 0.5em 1em 1em},clip]{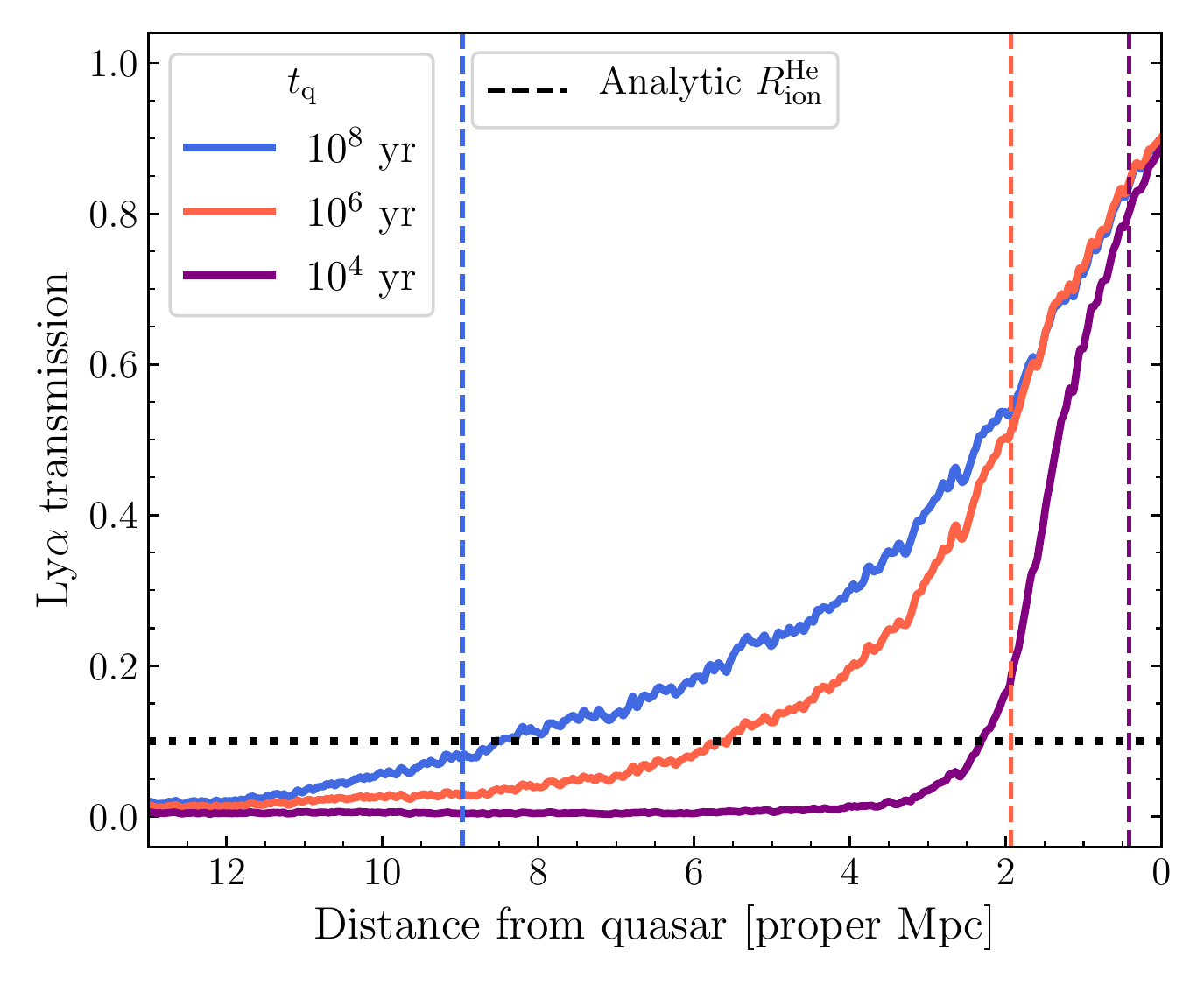}}
\end{center}
\caption{Stacked Ly$\alpha$ transmission for a $M_{1450}=-27$ quasar at $t_{\rm q}=10^4$ (purple), $10^6$ (red) and $10^8$ years (blue). While the early evolution of the proximity zone is dominated by non-equilibrium ionization of hydrogen, the additional transmission at late times ($t_{\rm q}>10^6$ yr) is due to heating from the reionization of He\,{\small II} by the quasar, where the correspondingly coloured vertical dashed lines show the progression of $R_{\rm ion}^{\rm He}$ following equation~(\ref{eqn:rion_he}).}
\label{fig:rt_stacks}
\end{figure}

We show the effect of this helium heating via stacked Ly$\alpha$ transmission spectra for a progression of $t_{\rm q}$ in Figure~\ref{fig:rt_stacks}. The evolution from $t_{\rm q}=10^4$ (purple) to $10^6$ years (red) is predominantly due to the non-equilibrium ionization of hydrogen at $10^4$ years discussed above -- after $10^6$ years, the hydrogen is in photoionization equilibrium, and so the profile should remain static at later times. However, we can see that after $10^8$ years (blue), there is substantially more transmission at very large radii. 
The impact of the helium heating can be seen by the extended transmission excess in the $10^8$ year stack relative to the $10^6$ year stack, where the dashed vertical lines show that the typical \ion{He}{3} region (computed via equation~\ref{eqn:rion_he}) increases in size from $\sim2$ to $\sim9$ proper Mpc
(the same effect has been seen in simulations of the He\,{\small II} proximity effect, \citealt{Khrykin16}). 

This ``helium bump" does not manifest as a sharp transition at $R_{\rm ion}^{\rm He}$ in the RT simulation stacks for a few different reasons. First, the inhomogeneous density field along the simulation skewers introduces scatter in the location of $R_{\rm ion}^{\rm He}$ from sightline to sightline. Second, the \ion{He}{3} ionization front itself is somewhat broad, on the scale of $\sim1$ proper Mpc (e.g. \citealt{Khrykin16}). Finally, there is considerable heating of the IGM beyond the \ion{He}{3} front from photons at X-ray energies ($\gtrsim500$ eV) that leads to an extended tail of heated gas at large radii (e.g. \citealt{McQuinn12,Davies16}).

\section{A New Semi-Analytic Model of Proximity Zones} \label{sec:sam}

It is worthwhile understanding where the $R_p$ scalings in the RT
simulations come from to build intuition into how to interpret the
observed proximity zones. 
In this section we build upon the analytic model of $R_p$ in the ionized IGM introduced by \citet{BH07} in two important ways. First, instead of assuming that the transmission through the IGM can be described by the GP optical depth, we calibrate a relation between the hydrogen photoionization rate and the \emph{effective} optical depth $\teff$ using our hydrodynamical simulation described in \S~\ref{sec:hydro}. Second, given the exceptionally small proximity zones in \citetalias{Eilers17} and the time dependencies seen in the RT simulations (\S~\ref{sec:rtrp}), we introduce non-equilibrium effects to explore how they could manifest in proximity zone measurements.

\subsection{Equilibrium model} \label{sec:eq}

A natural location one might predict for $R_p$ is the point where the
\emph{mean} transmission of the \lya forest in the enhanced ionizing
radiation field of the quasar is equal to 10\% \citep{Maselli09}. It
is not immediately obvious that this location is where the
\emph{first} crossing below 10\% transmission should be in individual
spectra, but as we will show later, it appears to be a reasonable
approximation. Similar to
\citet{Maselli09}\footnote{Contrary to \citet{Maselli09}, however,
  because we assume a post-reionization IGM we do not consider this
  $R_p$ to be a ``maximum" value, but rather a characteristic one.},
we define the mean transmission profile in terms of the
\emph{effective} optical depth $\teff\equiv -\ln{\langle F\rangle}$ as
determined from our hydrodynamical simulation instead of $\tau_{\rm
  GP}$, because $\teff$ explicitly accounts for the effect of density and
velocity structure of gas in the IGM on the Ly$\alpha$ transmission (on average).

\begin{figure}
\begin{center}
\resizebox{8.5cm}{!}{\includegraphics[trim={1em 1em 1em 1em},clip]{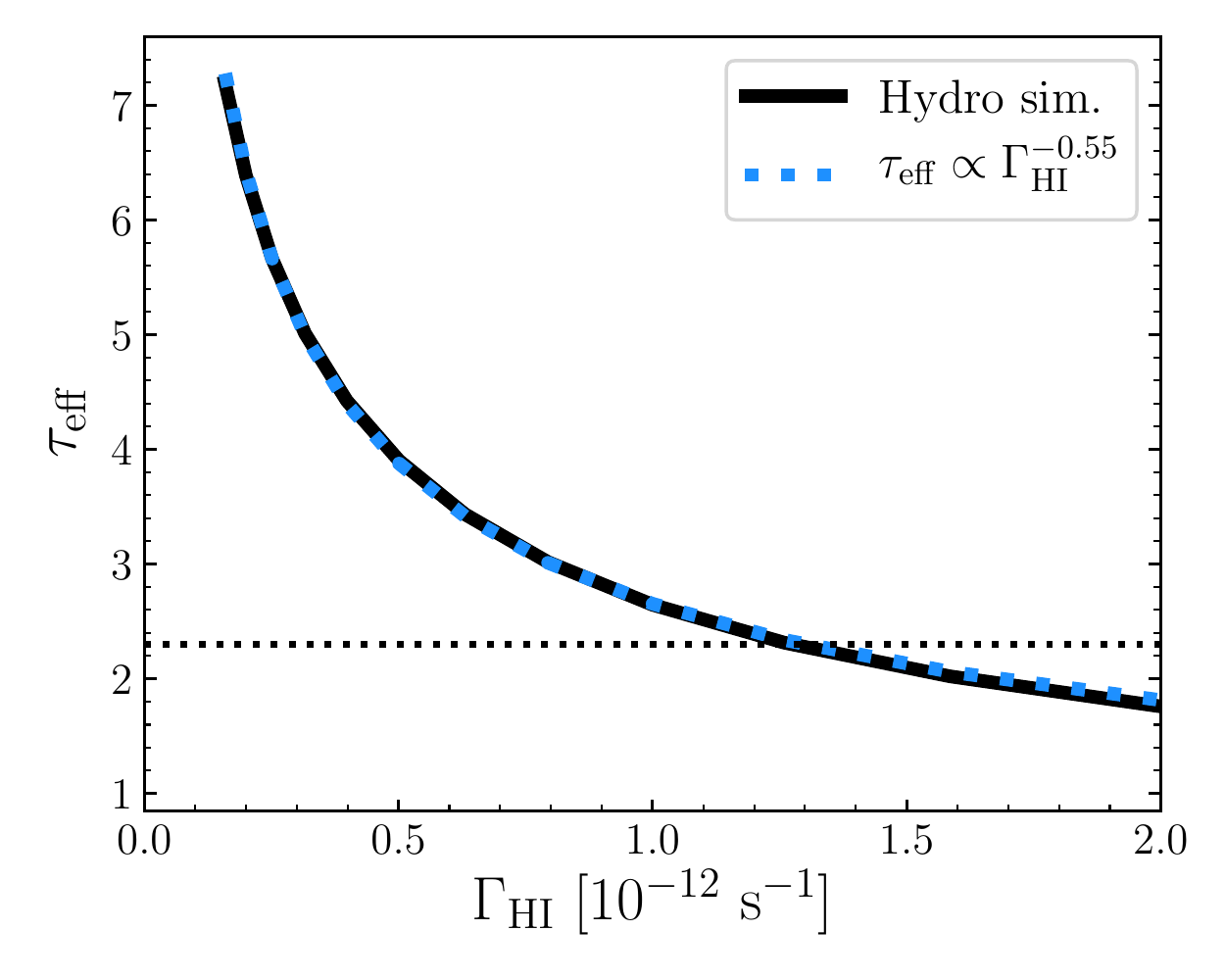}}
\end{center}
\caption{Relationship between $\tau_{\rm eff}$ and $\Gamma_{\rm HI}$ in our hydrodynamical simulation at $z=6$ (black curve) and a power law approximation $\teff\propto\Gamma_{\rm HI}^{-0.55}$ (blue dashed curve). The dotted curve shows $\teff=\tau_{\rm lim}=2.3$.}
\label{fig:teff_ghi}
\end{figure}

In Figure~\ref{fig:teff_ghi} we show the relationship between $\teff$
and the photoionization rate $\GHI$ measured by computing the mean
transmission through random skewers in our hydrodynamical simulation
with different $\GHI$. The dotted line shows the effective optical
depth corresponding to the 10\% transmission threshold, $\tau_{\rm
  lim}=-\ln(0.1)=2.3$. We see that $\teff=\tau_{\rm lim}$ at
$\GHI\approx 1.3\times10^{-12}$ s$^{-1}$. The behaviour of $\teff$ as a
function of $\GHI$ in the regime of interest ($\teff\sim2$--$6$) is
well-described by a power law, $\teff\propto\GHI^{-\alpha}$ with
$\alpha\approx0.55$\footnote{The exact fit is
$\teff = 5.678\,(\GHI/[2.5\times10^{-13}\,{\rm s}^{-1}])^{0.5486}$.}.
Note that this quantitative relationship between $\teff$ and $\Gamma_{\rm HI}$
is sensitive to the thermal state of the IGM in the simulation (e.g. \citealt{BB13}),
so the absolute scaling of Ly$\alpha$ transmission inside the proximity zone -- both in this analytic description
and in the radiative transfer simulations -- will vary depending on how the heat injection by reionization is modeled.
    
\begin{figure}
\begin{center}
\resizebox{8.5cm}{!}{\includegraphics[trim={1.5em 1em 1em 1em},clip]{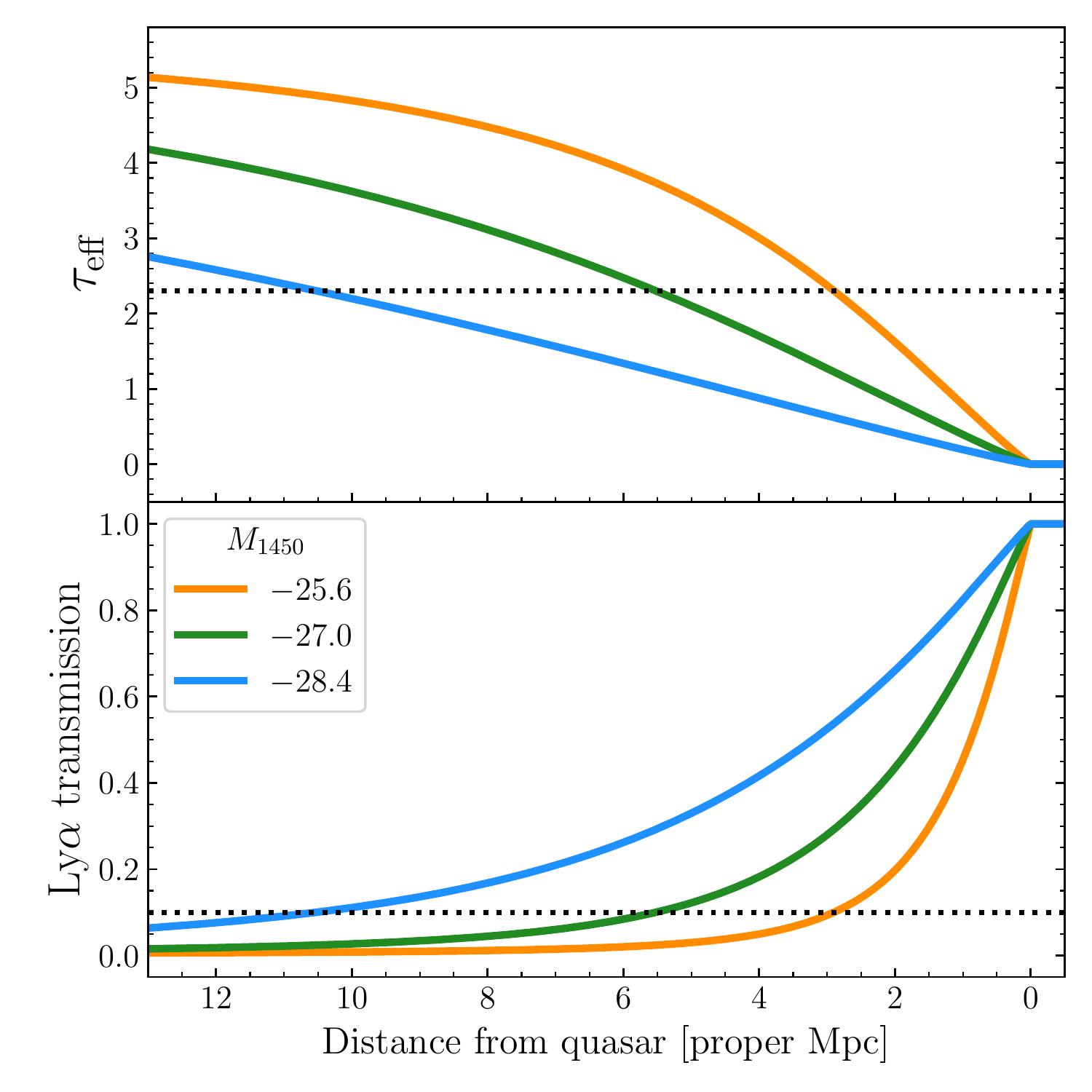}}
\end{center}
\caption{Equilibrium profiles of $\tau_{\rm eff}$ (top) and transmitted flux (bottom) for $M_{1450}$ = $-25.6$ (orange), $-27.0$ (green) and $-28.4$ (purple) computed via equation~(\ref{eqn:teffr}), compared to the $R_p$ threshold (dotted line).}
\label{fig:teff_r_eqmodel}
\end{figure}

Using this simple power-law relationship,
and neglecting the small-scale density enhancement close to the quasar
due to its massive host halo, we can estimate $\teff$ as a function of
line-of-sight distance $r$ from a quasar,
\begin{equation}
\teff(r) = \tau_{\rm b} \left( \frac{\Gamma_{\rm q}(r)+\Gamma_{\rm b}}{\Gamma_{\rm b}} \right)^{-\alpha},
\end{equation}
where $\Gamma_{\rm q}$ is the ionization rate from the quasar radiation alone, and $\tau_{\rm b}(\Gamma_{\rm b}, z)$ and $\Gamma_{\rm b}$ are the effective optical depth and ionization rate of the IGM in the absence of the quasar (i.e. the background opacity and ionization rate). 
Assuming that the mean free path of ionizing photons emitted by the quasar is large compared to our region of interest,
the ionization rate from the quasar can be written as $\Gamma_{\rm q}(r) = \Gamma_{\rm b} (r/r_{\rm b})^{-2}$ (see the lower panel of Figure~\ref{fig:rt_ex}),
where $r_{\rm b}$ is the distance at which the ionization rate from the quasar is equal to the average value in the ambient IGM.
For our assumed quasar SED, $r_{\rm b}$ is given by
\begin{eqnarray}
r_{\rm b} &=& 11.3 \left(\frac{\Gamma_{\rm b}}{2.5\times10^{-13}\,{\rm s}^{-1}}\right)^{-1/2} \nonumber \\
&& \times \left(\frac{\dot{N}_{\rm ion}}{1.73\times10^{57}\,{\rm s^{-1}}}\right)^{1/2}\ {\rm proper\ Mpc},
\end{eqnarray}
where $\dot{N}_{\rm ion}=1.73\times10^{57}$ s$^{-1}$ corresponds to a quasar with $M_{1450}=-27$ and $\Gamma_{\rm b}=2.5\times10^{-13}$ is consistent with constraints from the Ly$\alpha$ forest \citep{Davies17}.
The effective optical depth profile along the line of sight to the quasar can then be written as\footnote{\citet{Calverley11} derived a similar expression with $\alpha=1$ to describe the profile of the GP optical depth.}
\begin{equation} \label{eqn:teffr}
\teff(r) = \tau_{\rm b} \left[1+\left(\frac{r}{r_{\rm b}}\right)^{-2}\right]^{-\alpha}.
\end{equation}
We show example $\teff$ profiles for different quasar luminosities in Figure~\ref{fig:teff_r_eqmodel}. The proximity zone size $R_p$ is then found by solving for $r$ at the limiting optical depth $\tau_{\rm lim}$,
\begin{equation} \label{eqn:rp_eq}
R_p = r_{\rm b} \left[ \left( \frac{\tau_{\rm b}}{\tau_{\rm lim}} \right)^{1/\alpha}-1 \right]^{-1/2}.
\end{equation}

\begin{figure}
\begin{center}
\resizebox{8.5cm}{!}{\includegraphics[trim={1em 1.5em 1em 1em},clip]{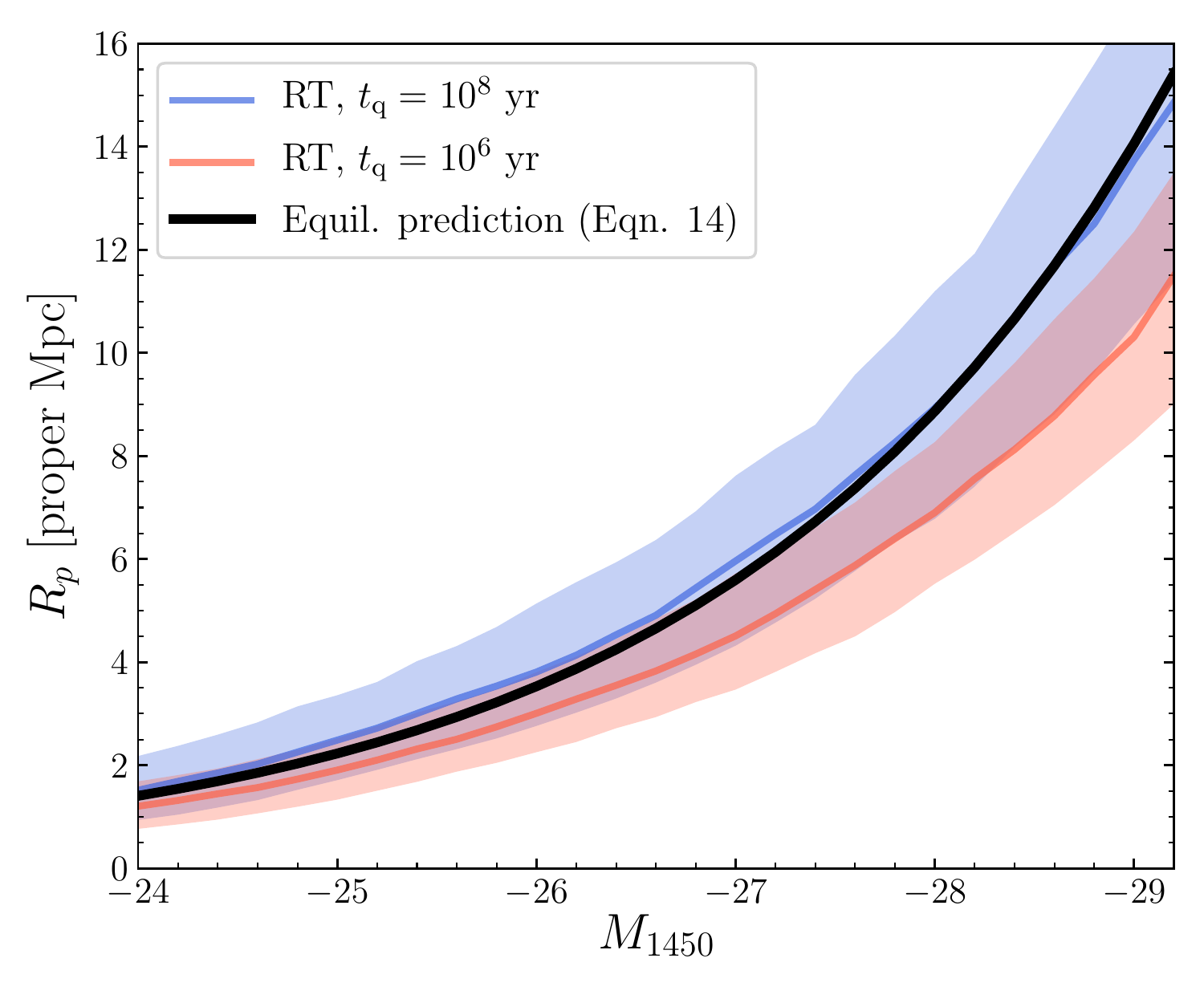}}
\end{center}
\caption{Comparison between our predicted equilibrium $R_p$ (black curve) and the output from the RT simulations at $t_{\rm q}=10^6$ years (red) and $10^8$ years (blue), where the curves and shaded regions indicate the median and 16--84th percentiles, respectively.}
\label{fig:teff_mag_eqmodel}
\end{figure}

From equation~(\ref{eqn:rp_eq}), we can derive the expected scalings of $R_p$ with various parameters. The scaling of $R_p$ with quasar ionizing photon output $\dot{N}_{\rm ion}$ is built into the definition of $r_{\rm b}$; at fixed $\Gamma_{\rm b}$, we have $r_{\rm b}\propto \dot{N}_{\rm ion}^{1/2}$, so we recover $R_p\propto \dot{N}_{\rm ion}^{1/2}$ as in BH07. In Figure~\ref{fig:teff_mag_eqmodel} we compare the predicted $R_p$ as a function of $M_{1450}$ (black) to the RT simulations at $10^6$ (red) and $10^8$ (blue) years. As mentioned previously, $\dot{N}_{\rm ion}^{1/2}$ is a slightly steeper dependence
than what we find in the RT simulations, due to the unaccounted-for effects of density fluctuations and He\,{\small II} reionization heating (\S~\ref{sec:helium}).

\begin{figure}	
 \begin{center}		
 \resizebox{8.5cm}{!}{\includegraphics[trim={1em 1em 1em 1em},clip]{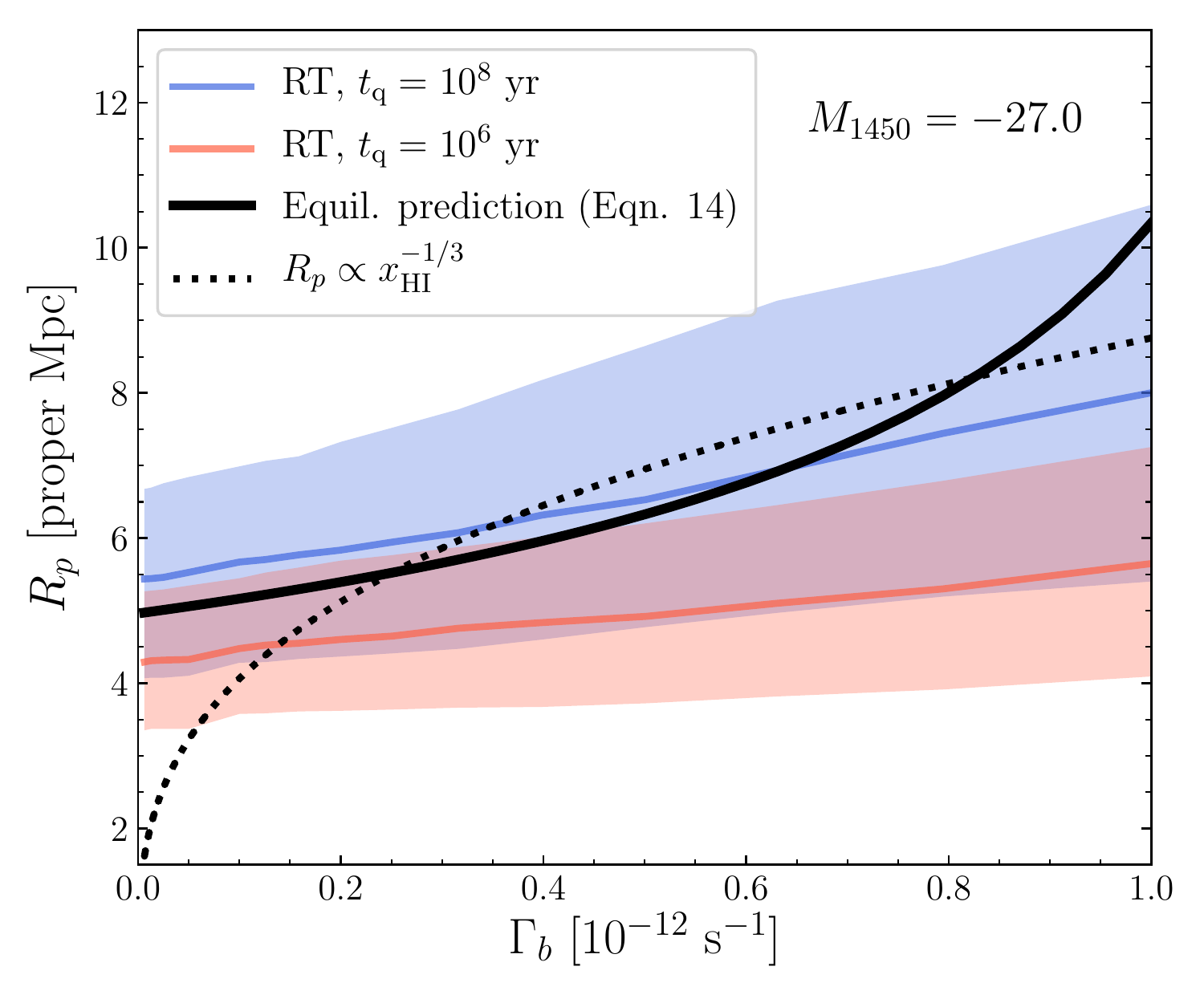}}		
 \end{center}		
 \caption{Dependence of $R_p$ on the background photoionization rate in the RT sims (coloured curves and shaded regions) compared to the equilibrium model prediction (black curve).}		
 \label{fig:rt_rp_uvb}		
 \end{figure}

Given the way we have written equation~(\ref{eqn:rp_eq}), the scaling with $\Gamma_{\rm b}$ is somewhat more complicated, as we must also account for the change in the background optical depth $\tau_{\rm b}\propto \Gamma_{\rm b}^{-\alpha}$. In the limit where the background opacity is much larger than the $R_p$ threshold, $(\tau_{\rm b}/\tau_{\rm lim})^{1/\alpha} \gg 1$, we have
\begin{equation}
R_p \approx r_{\rm b} (\tau_{\rm b}/\tau_{\rm lim})^{-1/2\alpha} \propto r_{\rm b} \Gamma_{\rm b}^{1/2} \propto \Gamma_{\rm b}^0,
\end{equation}
where in the last step we use the relation $r_{\rm b} \propto \Gamma_{\rm b}^{-1/2}$ at fixed $\dot{N}_{\rm ion}$ as mentioned above. Thus, if the IGM prior to the quasar turning on is highly opaque, $R_p$ should be insensitive to $\Gamma_{\rm b}$ \citepalias{Eilers17}.
As $\tau_{\rm b}$ approaches $\tau_{\rm lim}$, however, the predicted $R_p$ increases asymptotically,
whereas in the RT simulations, IGM density fluctuations can still cause the transmission to drop below 10\% even if the mean transmitted flux is greater than 10\%. In Figure~\ref{fig:rt_rp_uvb}, we compare the predicted $R_p$ vs $\Gamma_{\rm b}$ relationship from equation~(\ref{eqn:rp_eq}) to the RT simulations. The dependence of the analytic $R_p$ on $\Gamma_{\rm b}$ steepens greatly as $\tau_{\rm b}$ approaches $\tau_{\rm lim}$, while the RT simulations show a much weaker trend. We also show a dotted curve in Figure~\ref{fig:rt_rp_uvb} representing the interpretation of $R_p$ evolution in the context of $R_{\rm ion}$ by \citet{Fan06} and \citet{Carilli10} -- it is clear that this does not describe the dependence particularly well.

The evolution of $R_p$ with redshift depends on the cosmological and dynamical evolution of the IGM density field. We find that, between the different redshift outputs of our hydrodynamical simulation ($z=5.5,6.0,6.5$), $\tau_{\rm b}$ at fixed $\Gamma_{\rm b}$ evolves roughly as $(1+z)^{3.5}$. In the absence of $\Gamma_{\rm b}$ evolution, and assuming that $\tau_{\rm b} \gg \tau_{\rm lim}$, we then have $R_p\propto(1+z)^{-3.5/(2\alpha)}\sim(1+z)^{-3.2}$, practically identical to evolution measured from the RT simulations. This evolution is somewhat faster than the $\propto(1+z)^{2.25}$ relationship from \citet{BH07} (equation~\ref{eqn:rpbh07}) due to their neglect of the evolution of cosmological structure, and certainly faster than the $\propto(1+z)$ relationship predicted for a fully neutral IGM (derived from the $n_{\rm H}^{1/3}$ dependence in equation~\ref{eqn:rion}).

Let us now consider a fiducial $z=6$ quasar with $M_{1450}=-27$ in an IGM with $\Gamma_{\rm b}=2.5\times10^{-13}$ \citep{Davies17}, corresponding to $\tau_{\rm b}=5.7$ and $r_{\rm b}=11.3$ proper Mpc. From equation~(\ref{eqn:rp_eq}) we then expect $R_p = 5.5$ proper Mpc, similar to the observations and to the RT simulations with $t_{\rm q}=10^8$ years (see also Figure~\ref{fig:teff_mag_eqmodel}). However, the equilibrium model in equation~(\ref{eqn:rp_eq}) does not include a prescription for He\,{\small II} reionization heating, and thus it should instead be compared to the RT simulations with $t_{\rm q}\sim10^6$ years, which instead have $R_p = 4.5$ proper Mpc. Thus, the analytic approach moderately overestimates the normalization of $R_p$. This overprediction comes about due to the first-crossing definition of $R_p$ -- there is considerable path length at shorter distances where IGM fluctuations can drop the transmitted flux below 10\%. In the Appendix we further explore the effect of IGM fluctuations on proximity zone sizes, and show that including a prescription IGM fluctuations does indeed make the dependence on $\dot{N}_{\rm ion}$ shallower.

\subsection{Non-equilibrium model} \label{sec:noneq}		

The connection between $\teff$ and $\GHI$ assumed in
equation~(\ref{eqn:rp_eq}) relies on the assumption of ionization
equilibrium -- more accurately, the connection should be described in
terms of $\teff$ and $x_{\rm HI}\propto\GHI^{-1}$. When the quasar
turns on, $x_{\rm HI}$ in the vicinity of the quasar evolves over a
characteristic timescale $t_{\rm eq}=\GHI^{-1}$ (ignoring
recombinations). This behaviour is captured by the solution to
equation~(\ref{eqn:nhi}) assuming a constant ionization rate
(e.g. \citealt{McQuinn09b,Khrykin16}),
\begin{equation} \label{eqn:xhi}		
x_{\rm HI}(t) = x_{\rm HI,eq} + (x_{\rm HI,0}-x_{\rm HI,eq})e^{-t/t_{\rm eq}},		
\end{equation}		
where $x_{\rm HI,eq}$ and $x_{\rm HI,0}$ are the equilibrium
($t=\infty$) and initial ($t=0$) neutral fractions, respectively. We
can then re-write the above equation in terms of \emph{relative}
neutral fraction,
\begin{equation} \label{eqn:xhi_rel_evol}		
\frac{x_{\rm HI}(t)}{x_{\rm HI,0}} = 1+\left(\frac{x_{\rm HI,eq}}{x_{\rm HI,0}}-1\right)\left(1-e^{-t/t_{\rm eq}}\right).		
\end{equation}
We know that $\teff\propto\GHI^{-\alpha}$ and $x_{\rm HI,eq}\propto\GHI^{-1}$, so assuming that the IGM is initially in equilibrium with the UV background, we have		
\begin{equation}
\left(\frac{\teff(t)}{\tau_{\rm b}}\right)^{1/\alpha} = 1+\left(\frac{\Gamma_{\rm b}}{\GHI}-1\right)\left(1-e^{-t/t_{\rm eq}}\right).		
\end{equation}
Finally, substituting $t_{\rm eq}=1/\GHI$ and $\GHI=\Gamma_{\rm q}(r)+\Gamma_{\rm b}$, and assuming $\Gamma_{\rm HI,0}=\Gamma_{\rm b}$, we have
\begin{equation} \label{eqn:teffrt}
\teff(r,t) = \tau_{\rm b} \left[1-\left(\frac{\Gamma_{\rm q}(r)}{\Gamma_{\rm q}(r)+\Gamma_{\rm b}}\right)\left(1-e^{-t[\Gamma_{\rm q}(r)+\Gamma_{\rm b}]}\right)\right]^{\alpha},
\end{equation}
which asymptotes to equation~(\ref{eqn:teffr}) when $t \gg t_{\rm eq}$ as expected.

\begin{figure}
\begin{center}
\resizebox{8.5cm}{!}{\includegraphics[trim={2em 1em 1em 1em},clip]{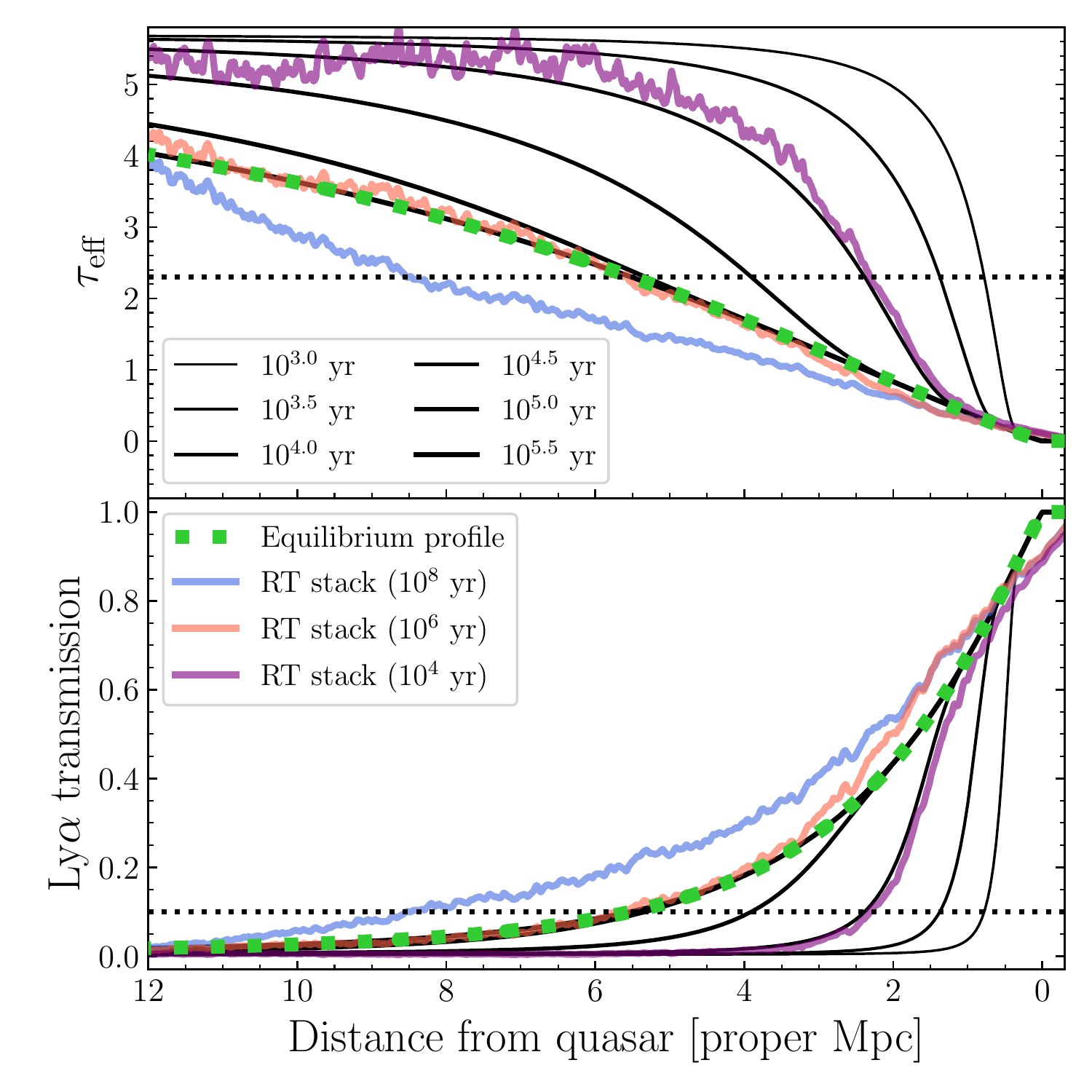}}
\end{center}
\caption{Non-equilibrium models of $\tau_{\rm eff}(r,t)$ (black curves) for a $M_{1450}$ quasar from equation~(\ref{eqn:teffrt}) for $t_{\rm q}\geq10^3$ yr (top panel) and the corresponding transmitted flux profiles (bottom panel). The stacked RT simulation spectra from Figure~\ref{fig:rt_stacks} are shown as coloured curves for comparison.}
\label{fig:teff_r_noneq}
\end{figure}

In Figure~\ref{fig:teff_r_noneq}, we show the $\teff(r,t)$ (top) and
$F=e^{-\teff}$ (bottom) profiles for a series of quasar ages, assuming
$M_{1450}=-27$ and $\Gamma_{\rm b} = 2.5\times10^{-13}$ s$^{-1}$. The
innermost regions equilibrate first, because $t_{\rm
  eq}\sim\Gamma_{\rm q}^{-1}\sim r^2$, and the entire proximity zone
(within the equilibrium $R_p$) has reached equilibrium within
$\sim10^5$ years. As discussed above in the context of the RT
simulations, any link between the small proximity zones seen in \citetalias{Eilers17}
and non-equilibrium effects thus requires quasar ages $\la10^5$
years. Figure~\ref{fig:teff_r_noneq} also shows the RT simulation stacks for $10^4$ (purple), $10^6$ (red) and $10^8$ years (blue). By comparing the RT simulation stacks to the analytic model, we can identify physical features that are unaccounted for in equation~(\ref{eqn:teffrt}).
At very small radii ($<1$ proper Mpc) the RT simulations show additional absorption due to dense gas associated with the massive quasar host halos (\S~\ref{sec:hydro}). On larger scales, we see deviations to low optical depth (high transmission) due to He\,{\small II} reionization heating, which propagates outward according to equation~(\ref{eqn:rion_he}) (see discussion in \S~\ref{sec:helium}). Importantly, at distances where we \emph{do} expect that the analytic model accounts for most of the physics, i.e. $\gtrsim1$ proper Mpc at $10^4$ years and $\gtrsim2$ proper Mpc at $10^6$ years, the agreement with the RT simulations is quite good.

\begin{figure}
\begin{center}
\resizebox{8.5cm}{!}{\includegraphics[trim={1em 1em 1em 1em},clip]{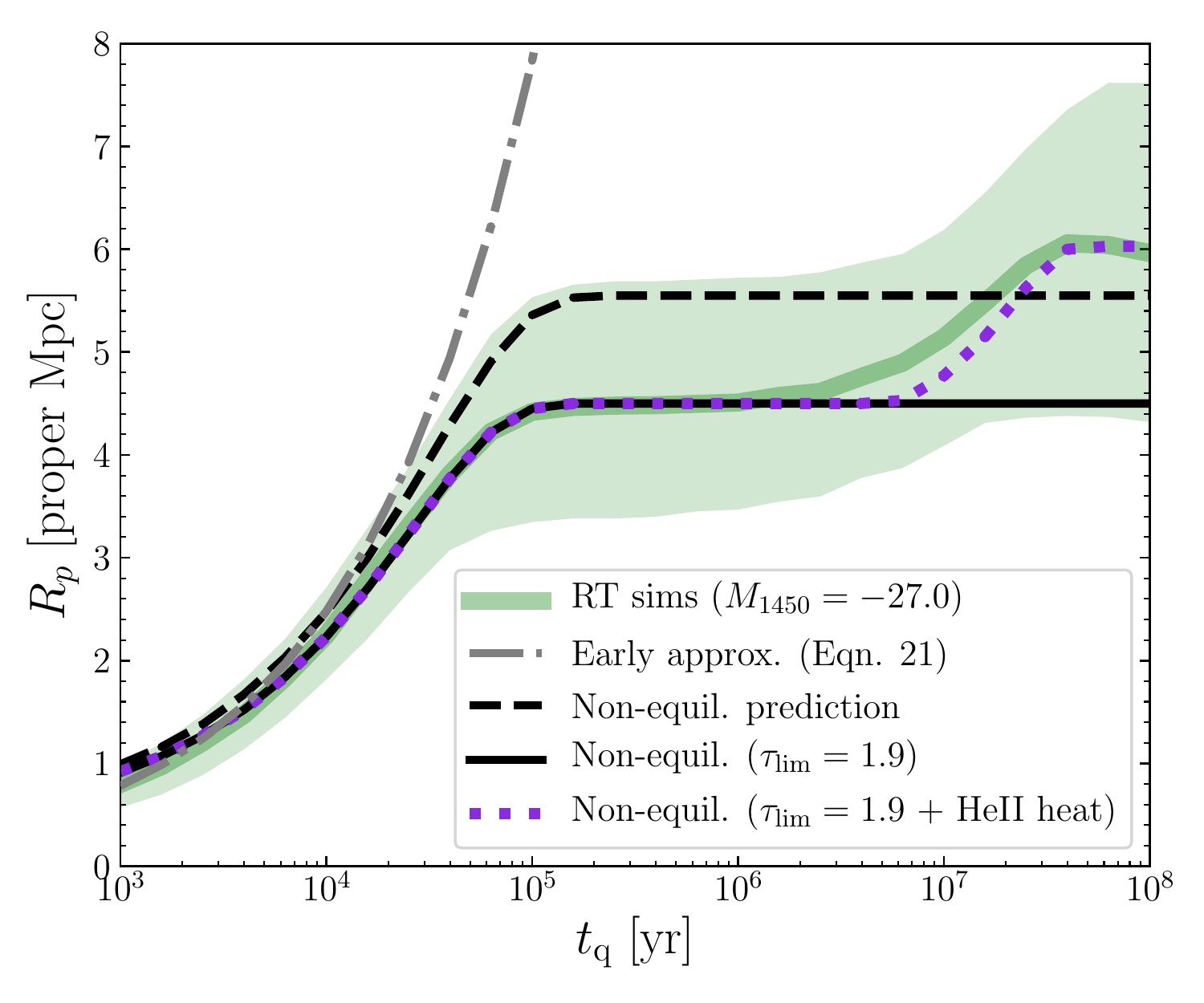}}
\end{center}
\caption{Non-equilibrium model for $R_p$ (black solid curve) compared to the RT simulations (green curve and shaded region). The dot-dashed grey curve shows the early-time approximation from equation~(\ref{eqn:early_rp}). The dashed black curve shows the model $R_p$ assuming $\tau_{\rm lim}=1.9$ to better match the $R_p$ normalization of the RT simulations. The dotted purple curve shows a toy model for the effect of He\,{\small II} reionization heating by the quasar.}
\label{fig:rp_tq_noneq}
\end{figure}

Equation~(\ref{eqn:teffrt}) does not have an analytic solution for $R_p$, although at radii where $\Gamma_{\rm q} \gg \Gamma_{\rm b}$ it simplifies to
\begin{equation}
\teff(r,t) \approx \tau_{\rm b} e^{-\alpha t\Gamma_{\rm q}(r)},
\end{equation}
from which we can then derive the following expression for the very early time evolution of $R_p$,
\begin{equation} \label{eqn:early_rp}
R_p \approx R_{p,{\rm eq}}\left(\frac{\alpha t_{\rm q} \Gamma_{\rm q}(R_{p,{\rm eq}})}{\ln{\tau_{\rm b}}-\ln{\tau_{\rm lim}}}\right)^{1/2},
\end{equation}
where $R_{p,{\rm eq}}$ is $R_p$ given by equation~(\ref{eqn:rp_eq}) and $\Gamma_{\rm q}(R_{p,{\rm eq}})\approx 10^{-12}$ s$^{-1}$. Thus at early times when $R_p$ is small, i.e. where $\Gamma_{\rm q}$ is very large, we have $R_p\propto t^{1/2}$. In Figure~\ref{fig:rp_tq_noneq} we compare this approximate behaviour, shown by the dot-dashed curve, to the RT simulations. While there is general agreement between the two for $t_{\rm q}\lesssim10^4$ yr, it is clear that this approximation fails spectacularly at all later times.

To better predict the evolution of $R_p$ on longer timescales, we measure $R_p$ from the full non-equilibrium $\teff(r,t)$ profiles (equation~\ref{eqn:teffrt}), as shown in Figure~\ref{fig:teff_r_noneq}, by numerically determining the distance where $\teff(r,t) = \tau_{\rm lim} = 2.3$
after smoothing the profiles with a 20\,{\AA} ($\approx1$ proper Mpc) top-hat filter.
The smoothing primarily affects the transmission
profile at very small radii $R\lesssim 1$ proper Mpc where the
proximity zone is ``contaminated" by the unabsorbed continuum of the
quasar redward of rest-frame Ly$\alpha$.
We show the resulting time evolution of $R_p$ as the dashed black curve in Figure~\ref{fig:rp_tq_noneq}. While the non-equilibrium model qualitatively matches
the early-time evolution of $R_p$, similar to equation~(\ref{eqn:rp_eq}) it consistently overpredicts $R_p$ compared to the RT simulations. As mentioned in the previous section, and explored further in the Appendix, this overprediction is due to the model not accounting for IGM density fluctuations, which can lead to fluctuations in transmission below 10\% when the mean transmission is higher than 10\%.
We find that a good match between the two at $t_{\rm q}<10^6$ yr can be obtained by instead measuring $R_p$ in the analytic model at $\tau_{\rm lim}=1.9$ ($F=0.15$). In the rest of this work we will apply this modified $\tau_{\rm lim}$ ``fudge factor" to predict the time-dependent behaviour of $R_p$ more accurately.

As noted in \S~2, the evolution of $R_p$ in the RT simulations shows a $\sim25$\% increase at $t_{\rm q}\sim10^{7-8}$ years due to the reionization of He\,{\small II} by the quasar. We can approximately include this effect in our analytic model by decreasing $\teff$ within the expected radius of the \ion{He}{3} bubble from equation~(\ref{eqn:rion_he}). As a toy model we approximate the amount of helium heating as a doubling of the IGM temperature (corresponding to $\sim10^4$ K of heat input, similar to expectations from RT simulations, e.g. \citealt{McQuinn09,Khrykin16}), and since $x_{\rm HI} \propto T^{-0.7}$, we have $\teff\propto T^{-0.7\alpha}$, so the resulting effect is a decrease in $\teff$ within $R_{\rm ion}^{\rm He}$ by a factor of $\approx 0.77$. The purple dotted curve in Figure~\ref{fig:rp_tq_noneq} shows the resulting ``helium bump," which roughly mimics the behaviour seen in the RT simulations. In detail, the effect of He\,{\small II} reionization heating in the RT simulations is somewhat more extended in time than predicted by this toy model due to the finite width of the ionization front and the extended X-ray heating at large distances discussed in \S~\ref{sec:helium}.

\section{Implications of non-equilibrium proximity zone behaviour} \label{sec:imp}

In the previous section, we derived an analytic approximation to the non-equilibrium behaviour of quasar proximity zones. In this section we use this new tool to show that
quasar light curves which vary on timescales comparable to $t_{\rm eq}$ can significantly modify the properties of quasar proximity zones.

\subsection{Blinking lightbulb model} \label{sec:blink}

In \S~\ref{sec:sam} we assumed that the quasar turns on once and has constant luminosity -- also known as the ``lightbulb" model for quasar light curves. However, quasars may instead undergo short periods of low and high accretion rates \citep{Novak11,Schawinski10,Schawinski15}, with flux-limited observations naturally drawing preferentially from the latter.
Here we use the analytic model to explore the effect of a simple ``blinking lightbulb" toy model where the quasar 
turns on and off with a fixed duty cycle\footnote{The ``duty cycle" of quasars is often defined in a cosmological context, i.e. as the fraction of the Hubble time that quasars are active. Here we use the term in a more simple manner to describe the fraction of time that the quasar is luminous after it initially turns on.}.
For simplicity we neglect heating due to He\,{\small II} reionization, which introduces time dependence on much longer timescales than the equilibration time\footnote{Any excursions to low $R_p$ will likely cause it to intersect with $R_{\rm ion}^{\rm He}$, even at relatively short quasar ages, so in principle the heating should always be included. However, here we wish to focus on the effects of quasar variability alone.}.

First let us consider what happens to the proximity zone when a quasar turns off (although, due to the absence of the quasar, it is not observable in this state). For a quasar turning off after an extended period of on-time (i.e. $t_{\rm on} \gg t_{\rm eq}$), the neutral fraction will relax from its equilibrium state as
\begin{equation}
\frac{x_{\rm HI}(t)}{x_{\rm HI,eq}^{\rm on}} = 1+\left(\frac{x_{\rm HI,b}}{x_{\rm HI,eq}^{\rm on}}-1\right)\left(1-e^{-t\Gamma_{\rm b}}\right),
\end{equation}
where $x_{\rm HI,b}$ is the equilibrium neutral fraction of the IGM in the absence of the quasar, and $x_{\rm HI,eq}^{\rm on}$ is the equilibrium neutral fraction when the quasar was turned on (i.e. the initial state of the gas in this case). The neutral fraction then returns to its original (quasar-free) state within a few $t_{\rm eq}\sim\Gamma_{\rm b}^{-1}$. Subsequent turn-on after an extended off-time $t_{\rm off}\gg t_{\rm eq}$ would then result in evolution identical to equation~(\ref{eqn:xhi}).

The situation is more complicated if the quasar is turning on and off on timescales comparable to $t_{\rm eq}$. In this case, the neutral fraction evolution can be solved as a chain of time-dependent ``on" and "off" solutions, with the initial conditions for one set by the final state of the other. In the ``on" state, the neutral fraction profile evolves according to equation~(\ref{eqn:xhi_rel_evol}),
\begin{equation}
\frac{x_{\rm HI}(r,t)}{x_{\rm HI}(r,t_{\rm on})} =  1+\left(\frac{x_{\rm HI,eq}^{\rm on}(r)}{x_{\rm HI}(r,t_{\rm on})}-1\right)\left(1-e^{-(t-t_{\rm on})(\Gamma_{\rm q}(r)+\Gamma_{\rm b})}\right),
\end{equation}
where $t_{\rm on}$ is the time at which the quasar turned on. In the ``off" state, we have instead
\begin{equation}
\frac{x_{\rm HI}(r,t)}{x_{\rm HI}(r,t_{\rm off})} = 1+\left(\frac{x_{\rm HI,b}}{x_{\rm HI}(r,t_{\rm off})}-1\right)\left(1-e^{-(t-t_{\rm off})\Gamma_{\rm b}}\right),
\end{equation}
where $t_{\rm off}$ is the time at which the quasar turned off. The effective optical depth profiles, and $R_p$, can then be computed via the procedure described in \S~\ref{sec:noneq}.

\begin{figure}
\begin{center}
\resizebox{8.5cm}{!}{\includegraphics[trim={1em 1em 1em 1em},clip]{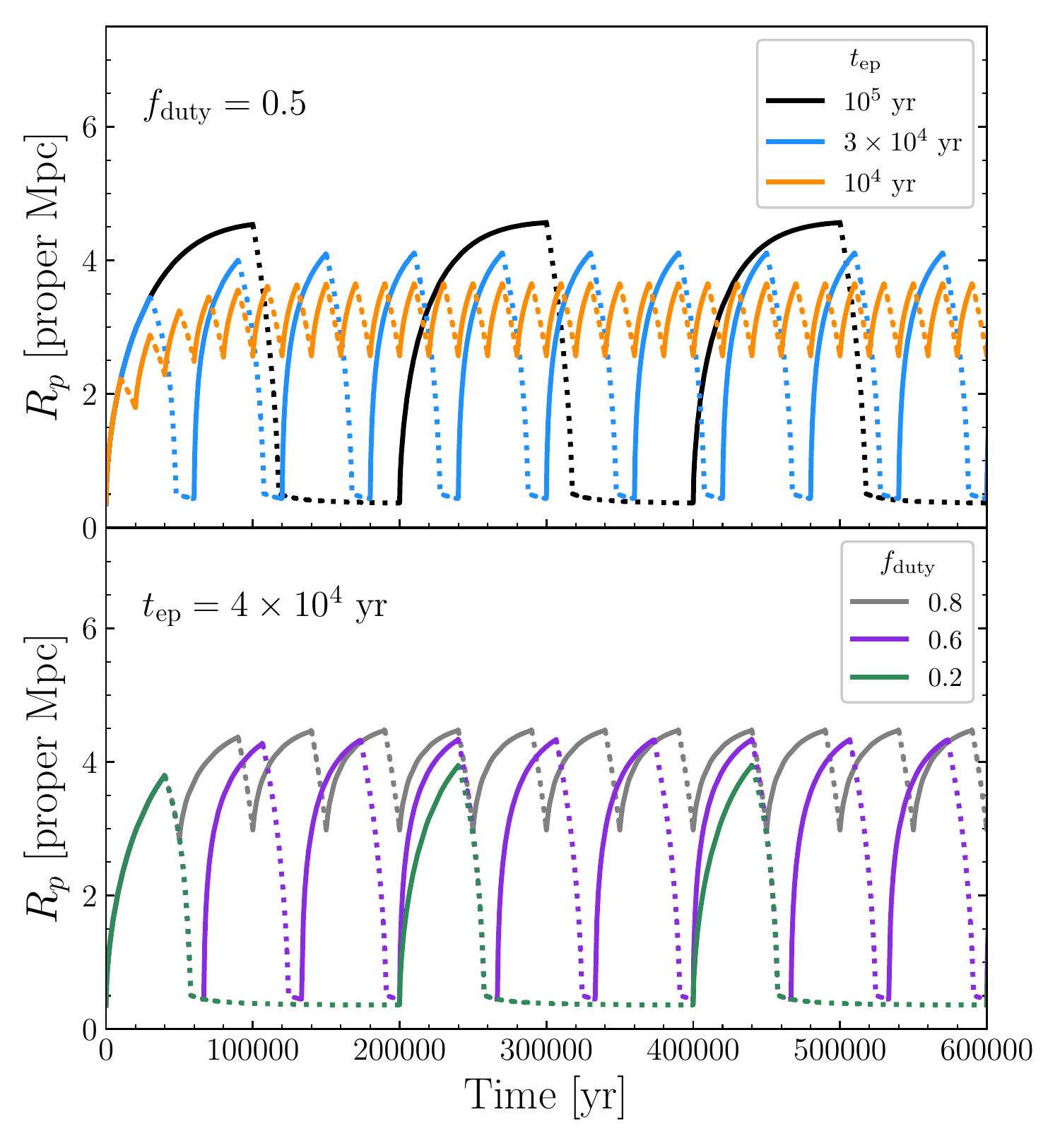}}
\end{center}
\caption{Analytic $R_p$ evolution for ``blinking lightbulb" light curve models. Solid curves show the evolution while the quasar is on (i.e. observable), while the dotted curves show the evolution when the quasar is off (i.e. invisible). Top: Varying episodic lifetime at a fixed duty cycle $f_{\rm duty}=0.5$. Bottom: Varying duty cycle with a fixed episodic lifetime $t_{\rm ep}=4\times10^4$ years.}
\label{fig:flicker}
\end{figure}

\begin{figure}
\begin{center}
\resizebox{8.5cm}{!}{\includegraphics[trim={1em 1em 1em 1em},clip]{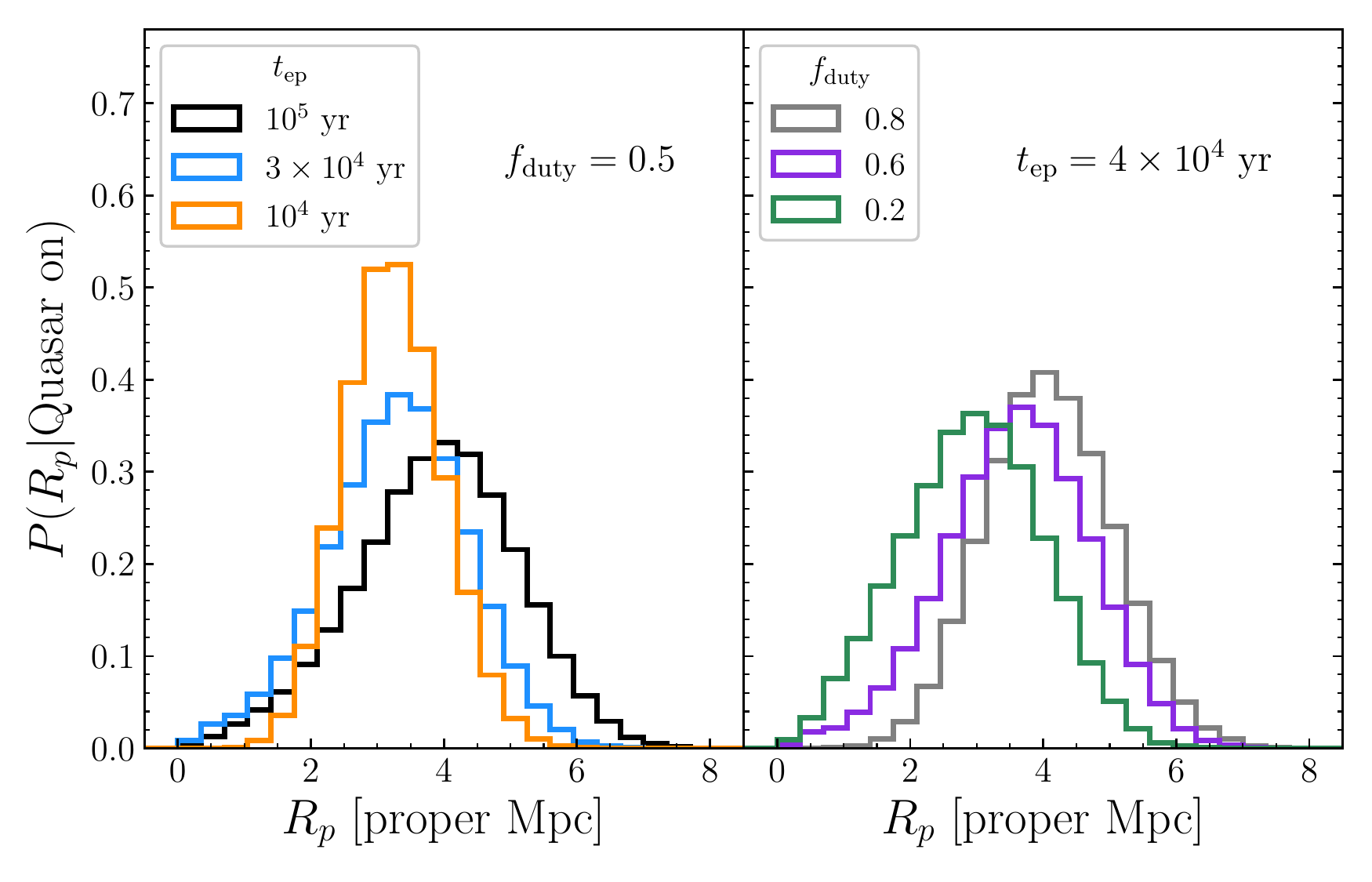}}
\end{center}
\caption{Distributions of $R_p$ in the analytic ``blinking lightbulb" light curve models shown in Figure~\ref{fig:flicker}. Scatter similar to the RT simulations has been added to realistically broaden the distributions. Left: Varying episodic lifetime at a fixed duty cycle $f_{\rm duty}=0.5$. Right: Varying duty cycle with a fixed episodic lifetime $t_{\rm ep}=4\times10^4$ years.}
\label{fig:flicker_hist}
\end{figure}

In the top panel of Figure~\ref{fig:flicker} we show examples of $R_p$
evolution computed in this fashion for such ``blinking lightbulb" quasars with duty cycle
$f_{\rm duty}=0.5$ and varying episodic lifetime $t_{\rm ep}$, where the ``on"
state (when the quasar can be observed) is shown by solid curves and
the ``off" state (when the quasar is invisible) is shown by dotted
curves. For episodes (and off times) longer than the equilibration time, shown by the black curve, we see roughly identical behaviour to the original lightbulb case. The blue curve shows that as the episodes get shorter, the proximity zone never has enough time to fully equilibrate, and so the maximum $R_p$ starts to become truncated. If the episodic lifetime is substantially smaller than $t_{\rm eq,off}$
(i.e. the orange curve), the region of enhanced transmission does not
have time to relax completely, and $R_p$ oscillates around a value
corresponding to the equilibrium size for $\dot{N}_{\rm ion} \times f_{\rm
  duty}$. In the bottom panel of Figure~\ref{fig:flicker} we show
quasars with fixed episodic lifetime $t_{\rm ep}=4\times10^4$ yr and
varying duty cycle. Large duty cycles, shown by the grey curve, result in a narrow range of possible $R_p$ values, in particular not allowing $R_p$ to be very small. Smaller duty cycles, shown by the purple and green curves, regularly return to small $R_p$ as the IGM is allowed to return closer to its original ionization state.

If observations of quasar proximity zones sample such flickering
lightcurves, then the observed distribution of $R_p$ will reflect
samples from the ``on" curves (solid) in Figure~\ref{fig:flicker} plus
scatter
due to IGM fluctuations.
We approximate IGM fluctuations by assuming that $R_p$ is drawn from a Gaussian centered on the typical value with scatter that varies as $\sigma(R_p)=A+BR_p^C$,  where we fit $A=0.155$, $B=0.091$, and $C=1.478$ to the scatter in the RT simulations at $t_{\rm q}<2\times10^5$ years for a $M_{1450}=-27$ quasar.
In Figure~\ref{fig:flicker_hist} we show
samples from the varying episodic lifetime (left panel) and varying duty cycle
(right panel) lightcurves shown in Figure~\ref{fig:flicker}. Existing
samples of quasar spectra should be able to distinguish between these
distributions, and thus constrain both the lifetime and duty cycle of
luminous quasar activity.

\subsection{General lightcurves} \label{sec:matt}

For more complicated quasar lightcurves, equation~(\ref{eqn:nhi}) can be solved more generally,
\begin{equation} \label{eqn:general}
x_{\rm HI}(t) = x_{\rm HI,0} e^{-\int_0^t \Gamma(t') dt'} + \alpha_\HII n_e \int_0^t e^{-\int_{t'}^t \Gamma(t'') dt''} dt',
\end{equation}
where $\Gamma(t)$ is the total H\,{\small I} ionization rate. We can then write the non-equilibrium effective optical depth profile as
\begin{eqnarray} \label{eqn:taugen}
\tau_{\rm eff}(r,t) = \tau_{\rm eff}(r,0)\Big[e^{-\int_0^t \Gamma_{\rm HI}(r,t') dt'} \\ \nonumber +\Gamma_{\rm HI}(r,0)\int_0^t e^{-\int_{t'}^{t}\Gamma_{\rm HI}(r,t'') dt''} dt'\Big]^\alpha,
\end{eqnarray}
where $\Gamma_{\rm HI}(r,t)$ is the background ionization rate plus the variable contribution from the quasar, and we have assumed ionization equilibrium at $t=0$, i.e. $x_{\rm HI,0} = \alpha_\HII n_e/\Gamma_{\rm HI}(r,0)$.

\begin{figure}
\begin{center}
\resizebox{8.5cm}{!}{\includegraphics[trim={1.1em 1em 1em 1em},clip]{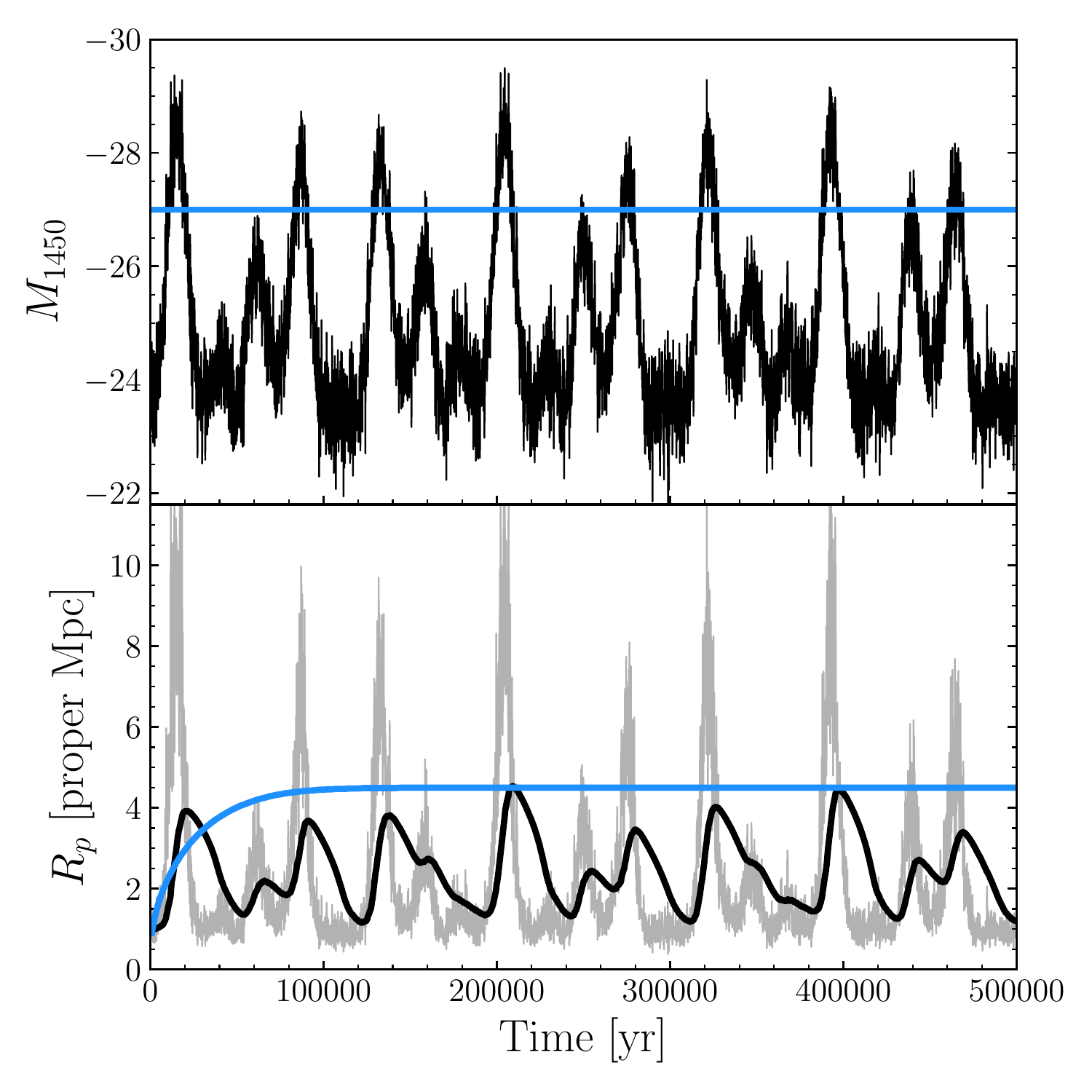}}
\end{center}
\caption{Top: Toy quasar lightcurve with short and long timescale variability (black) and a typical lightbulb model (blue). Bottom: Resulting $R_p$ evolution computed via equation~(\ref{eqn:taugen}) for the variable lightcurve (black) and lightbulb (blue) models. The grey curve shows the equilibrium $R_p$ of the variable lightcurve at each instantaneous luminosity computed via equation~(\ref{eqn:rp_eq}).}
\label{fig:generic}
\end{figure}

\begin{figure}
\begin{center}
\resizebox{8.5cm}{!}{\includegraphics[trim={1.0em 1em 1em 1em},clip]{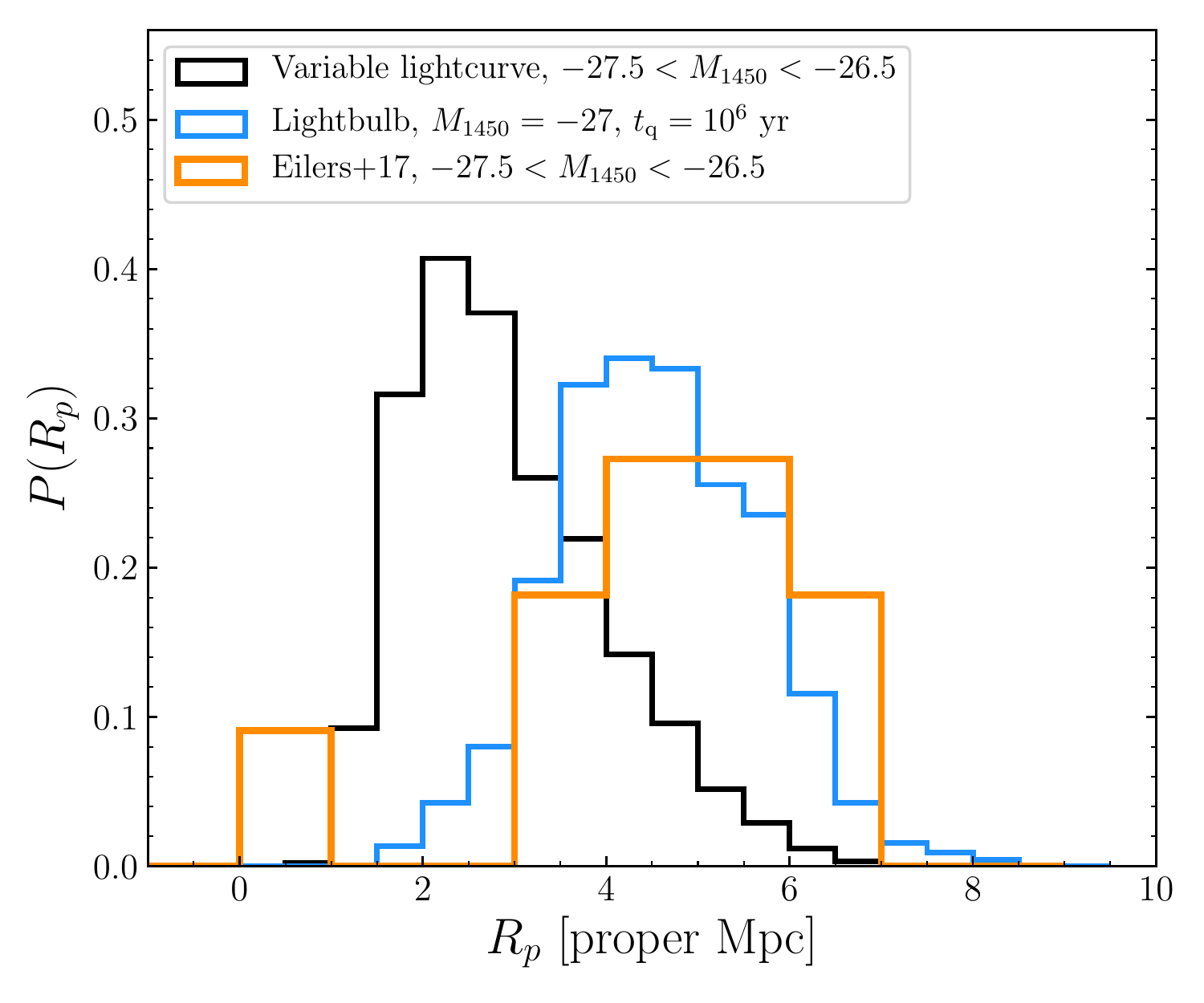}}
\end{center}
\caption{The distributions of $R_p$ determined by sampling the variable lightcurve from Figure~\ref{fig:generic} when the quasar has $-27.5 < M_{1450} < -26.5$ (black), the RT simulations assuming a lightbulb model with $M_{1450}=-27$ at $t_{\rm q}=10^6$ years (blue), and observed $z\sim6$ quasars with $-27.5 < M_{1450} < -26.5$ and accurate systemic redshifts.}
\label{fig:generic_hist}
\end{figure}

In Figure~\ref{fig:generic} we show a toy quasar lightcurve and its corresponding $R_p$ evolution by computing $\tau_{\rm eff}(r,t)$ profiles following equation~(\ref{eqn:taugen}) and numerically determining $R_p$.
The toy lightcurve,
shown in the top panel, consists of short timescale ($\Delta t\sim100$ yr) and long timescale ($\Delta t\sim10^4$ yr) variations over multiple orders of magnitude in luminosity. The duration of the long timescale variations was chosen to be comparable to the equilibration timescale at the equilibrium $R_p$, i.e. $\sim1/\Gamma_{\rm HI}(R_{p,{\rm eq}})\sim2.5\times10^4$ yr.
The resulting $R_p$ evolution, shown by the black curve in the bottom panel, reflects mostly the long
timescale evolution, as the ionization equilibration time is $\sim1/\Gamma_{\rm HI}(R_p)\sim2.5\times10^4$ yr. This strong variability decouples the observed luminosity of the quasar from $R_p$, as demonstrated by the substantial disagreement with the grey curve in the bottom panel which shows the predicted equilibrium $R_p$ from equation~(\ref{eqn:rp_eq}). The blue curve in the bottom panel of Figure~\ref{fig:generic} also shows the evolution of a lightbulb lightcurve quasar with $M_{1450}=-27$ for comparison. Despite the variable lightcurve frequently reaching brighter magnitudes, $R_p$ fails to reach the equilibrium size for the lightbulb case.

A variable lightcurve, such as the toy model presented in Figure~\ref{fig:generic}, will thus affect the distribution of observed $R_p$ at a fixed $M_{1450}$. The black curve in Figure~\ref{fig:generic_hist} shows the resulting distribution of $R_p$ when observed during times when $-27.5 < M_{1450} < -26.5$, with approximate IGM density fluctuations added as in \S~\ref{sec:blink}. The distribution is strongly skewed towards small $R_p$, with only rare excursions to $R_p>5$ proper Mpc. The blue curve shows the corresponding distribution of $R_p$ for the lightbulb model, drawn from the RT simulations (\S~\ref{sec:rt}). The orange curve shows the observed distribution of $R_p$ from \citetalias{Eilers17} for $z\sim6$ quasars with $-27.5 < M_{1450} < -26.5$ and systemic redshifts coming from either sub-millimeter observations or the \ion{Mg}{2} emission line (i.e., accurate enough that their uncertainties are subdominant compared to IGM fluctuations). The observed $R_p$ distribution is clearly incompatible with the toy lightcurve model, but consistent with the $10^6$ year lightbulb model (and as shown in \citetalias{Eilers17}, a $10^{7.5}$ year lightbulb model is consistent with the observations as well).

\section{Discussion \& Conclusion}

In this work we presented a suite of RT simulations of quasar proximity zones at $z=6$ to investigate the time-dependent properties of the effective proximity zone size $R_p$.
The early time ($t_{\rm q}\lesssim10^5$ yr) behaviour of $R_p$ is dominated by equilibration of the IGM to the newly elevated ionization rate close to the quasar (\citetalias{Eilers17}; \citealt{Eilers18J1335}), while the late time ($t_{\rm q}\gtrsim10^7$ yr) behaviour of $R_p$ demonstrates a sensitivity to the thermal proximity effect from He\,{\small II} reionization. Leveraging the connection between the ionization rate and effective optical depth of the Ly$\alpha$ forest from a hydrodynamical simulation, we constructed a novel semi-analytic model for the proximity zone to better understand its dependencies on quasar age, luminosity, UV background, and redshift. By basing our model on the \emph{effective} optical depth of the IGM rather than the GP optical depth, our model provides a quantitative picture for $R_p$ that only suffers from the lack of IGM density fluctuations, which we explored in the Appendix. Our model also considers non-equilibrium ionization inside post-reionization proximity zones for the first time\footnote{The first time for \emph{hydrogen}, that is, see \citet{Khrykin19} for discussion of the much longer equilibration timescale for He\,{\small II} proximity zones at $z\lesssim4$.}, in light of the discovery of apparently-young quasars with exceptionally small proximity zones (\citetalias{Eilers17}; \citealt{Eilers18J1335}). Finally, we employed this model as a tool to efficiently explore how measurements of proximity zone properties can reflect quasar variability on timescales comparable to the equilibration time, or even longer if the ``helium bump" can be detected.

Our predictions for the effect of quasar variability on $R_p$ in \S~\ref{sec:imp} suggest that the ensemble properties of high-redshift quasar proximity zones can be used to constrain the lifetime and duty cycle of luminous quasar events. Generically, strong quasar variability on timescales comparable to $t_{\rm eq} \sim 1/\Gamma_{\rm HI}(R_p) \sim 2.5\times10^4$ yr can ``decouple" the present-day observed quasar luminosity from the size of the proximity zone as measured by $R_p$, as shown by the discrepancy between the black and grey curves in Figure~\ref{fig:generic}.
The good agreement between the majority of $R_p$ measurements at $z\sim6$ and the model predictions for $t_{\rm q}\sim10^{6-8}$ years \citepalias{Eilers17} should rule out such strong variability -- however, the existence of a handful of quasars with very small $R_p$ (\citetalias{Eilers17}; \citealt{Eilers18J1335}) suggests that strong variability on timescales $\lesssim10^4$ years does in fact exist, but must be relatively rare.

We note that the quantitative statements in this work regarding the absolute scale of $R_p$ as a function of observed quasar properties depend on our choice of hydrodynamical simulation in two ways. First, as mentioned in \S~\ref{sec:sam} the state of the hydrodynamical simulation at $z\sim6$ reflects the assumption of a uniform reionization event at $z>10$ with minimal heat injection, which is inconsistent with observations of reionization occurring at much lower redshift (e.g. \citealt{Planck18,Davies18b}) and predictions of inhomogeneous heating of the IGM by $\sim2\times10^4$ K (e.g. \citealt{D'Aloisio18b,Onorbe18}). In future work we will investigate how quasar proximity zones in a more realistic hydrodynamical simulation incorporating these features differs from the model presented here. Second, given the small volume of our simulation $(100\,\rm{Mpc}/h)^3$ relative to the $\sim1\,\rm{Gpc}^{-3}$ space density of luminous $z\sim6$ quasars (e.g. \citealt{Jiang16}), we may be underestimating the effect of the local quasar environment on the structure of the proximity zone, although the dependence with halo mass has been shown to be fairly weak \citep{BH07,Keating15}.

Future samples of high-quality $z\sim6$ quasar spectra with precise systemic redshift estimates from current observational programs will soon increase the number of $z\sim6$ proximity zone measurements by a factor of a few. Studying these proximity zones in the context of the non-equilibrium phenomenology developed here will allow for powerful constraints on the lightcurves of accreting supermassive black holes, especially when combined with additional IGM-based probes of quasar activity at lower redshift that are sensitive to much longer timescales 
\citep{Khrykin16,Khrykin17,Khrykin19,Schmidt17,Schmidt18,Schmidt18echo}. 
While we have focused on $R_p$ in this work, in principle the timescale at which the proximity zone is sensitive to variability differs along the line of sight as $t_{\rm eq}\propto r^2$. By investigating the entire transmission profile of an ensemble of $z\sim6$ quasars, it should then be possible to constrain the power spectrum of quasar variability across a wide range of timescales that will not be directly observable for millennia.

\section*{Acknowledgements}

We thank Z. Luki\'{c} for providing access to the Nyx hydrodynamical simulation, and we thank M. McQuinn for helpful discussions during which he derived equation~(\ref{eqn:general}) for us. FBD acknowledges support from the Space Telescope Science Institute, which is operated by AURA for NASA, through the grant \emph{HST}-AR-15014.

%%%%%%%%%%%%%%%%%%%%%%%%%%%%%%%%%%%%%%%%%%%%%%%%%%


\bibliographystyle{mnras}
 \newcommand{\noop}[1]{}

%%%%%%%%%%%%%%%%%%%%%%%%%%%%%%%%%%%%%%%%%%%%%%%%%%

%%%%%%%%%%%%%%%%% APPENDICES %%%%%%%%%%%%%%%%%%%%%

\appendix

\section{Analytic model for scatter in proximity zone sizes} \label{sec:scatter}

The analytic models discussed in \S~\ref{sec:sam} make predictions for $R_p$ from the mean transmission profiles in the proximity zone. In reality, however, there is significant scatter due to IGM density fluctuations, and $R_p$ is defined as the \emph{first} time that the transmission falls below 10\%. Here we introduce a method to estimate the effect of IGM scatter on proximity zone sizes -- however, for simplicity, we assume that the observed spectrum is additionally \emph{re-binned} to the size of the smoothing scale\footnote{This alternative definition already exists in the literature, e.g. \citet{Willott07}, but most works are not clear about whether their smoothing procedure also includes re-binning of the spectrum.}, 20\,{\AA} in the observed frame or $\approx1$ proper Mpc at $z=6$, to avoid having to model strong correlations from pixel to pixel. We denote this alternative $R_p$ definition as $R_p^{\rm alt}$.

Assuming no correlations between pixels, the probability for $R_p^{\rm alt} = r_i$ for any (binned) pixel $i$ can be written as
\begin{equation} \label{eqn:rpflucts}
P(R_p^{\rm alt}=r_i) = \left(\prod_{r_j < r_i} P(\tau_{{\rm eff},j} < \tau_{\rm lim})\right) \times P(\tau_{{\rm eff},i} > \tau_{\rm lim}),
\end{equation}
where $\tau_{{\rm eff},j}$ is the effective optical depth for pixel $j$ at distance $r_j$. That is, the probability that any given distance $r_i$ is equal to $R_p^{\rm alt}$ is the probability that all smaller radii have $\tau_{\rm eff} < \tau_{\rm lim}$ times the probability that $\tau_{\rm eff}(r_i) > \tau_{\rm lim}$.

We estimate $P(\tau_{\rm eff})$ by calibrating the strength of IGM density fluctuations on the smoothing scale from our Nyx hydrodynamical simulation, similar to the prescription for $\teff(\Gamma_{\rm HI})$ calibrated in \S~\ref{sec:eq}. We find that the distribution of $\teff$ on 1 proper Mpc scales, when the mean transmission is 10\%, very closely follows a lognormal distribution (e.g. \citealt{Becker07}) with $\sigma_{\rm ln \tau}\approx0.3$. We can then rewrite equation~(\ref{eqn:rpflucts}) as
\begin{eqnarray} \label{eqn:rpflucts2}
P(R_p^{\rm alt}=r_i) &=& \left(\prod_{r_j < r_i} \frac{1}{2}\left[1-{\rm erfc}\left(\frac{\ln{\bar{\tau}_{\rm eff}(r_j)}-\ln{\tau_{\rm lim}}}{\sqrt{2} \sigma_{\rm ln \tau}}\right)\right]\right) \nonumber \\
&\times& \frac{1}{2} {\rm erfc}\left(\frac{\ln{\bar{\tau}_{\rm eff}(r_i)}-\ln{\tau_{\rm lim}}}{\sqrt{2} \sigma_{\rm ln \tau}}\right),
\end{eqnarray}
where $\bar{\tau}_{{\rm eff},j}$ is the mean effective optical depth expected in pixel $j$ (i.e. from the mean transmitted flux profile as described in the previous sections). Note that equation~(\ref{eqn:rpflucts2}) is a very rough approximation that does not take into account correlations in the IGM density field or the gradient in $\tau_{\rm eff}$ inside of each 1 proper Mpc pixel. Nevertheless, it should provide a baseline expectation for the fluctuations in $R_p^{\rm alt}$.

\begin{figure}
\begin{center}
\resizebox{8.5cm}{!}{\includegraphics[trim={0 0 0 0},clip]{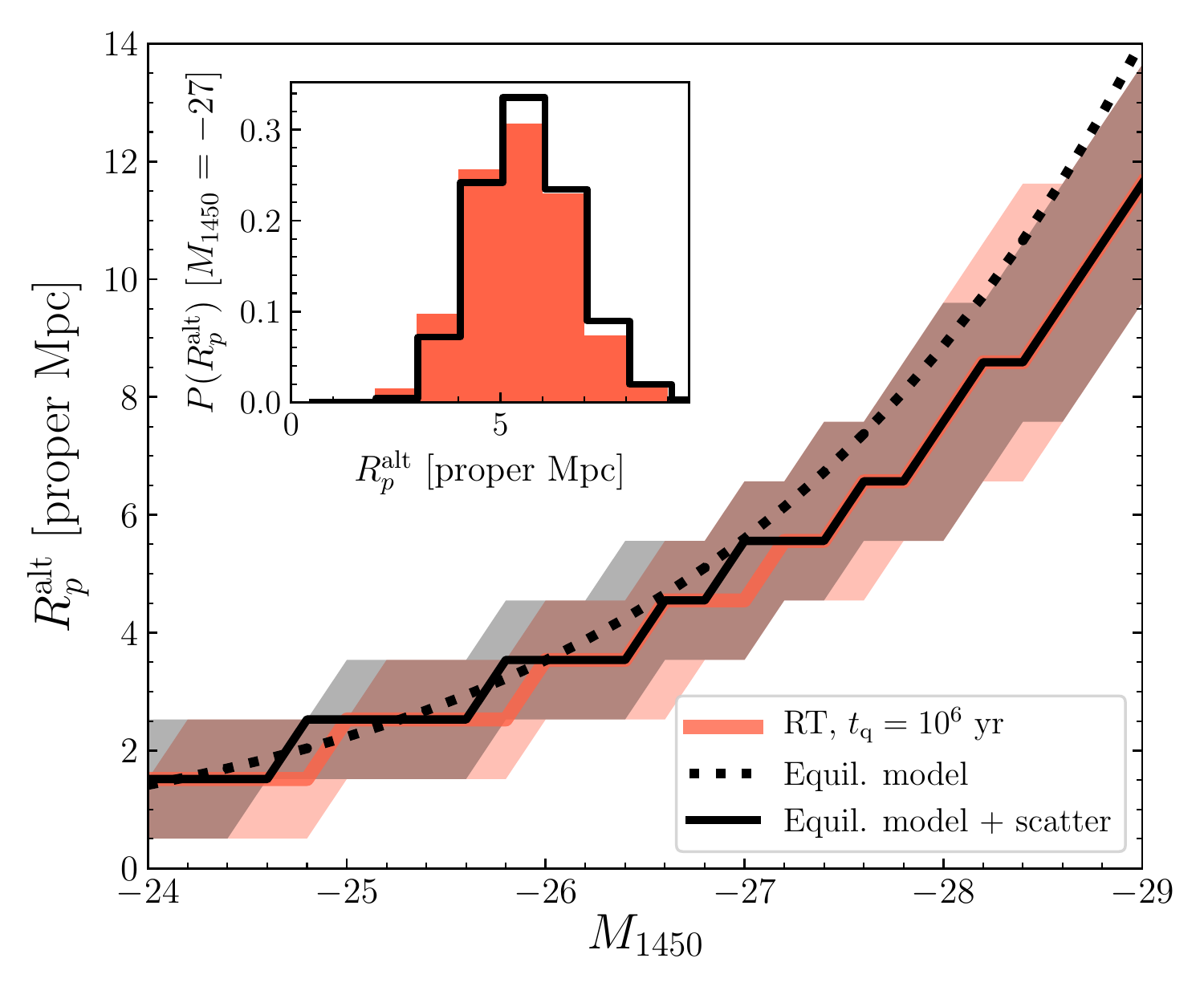}}
\end{center}
\caption{Equilibrium model for $R_p^{\rm alt}$ with approximate scatter (black, equation~\ref{eqn:rpflucts2}) compared to the RT simulations at $t_{\rm q}=10^6$ years (red) and the equilibrium model for $R_p$ without scatter from the main text (dotted black, equation~\ref{eqn:rp_eq}). Solid curves show median values, while the shaded regions indicate the 16--84th percentile scatter. The inset panel shows the respective distributions of $R_p^{\rm alt}$ at $M_{1450}=-27$.}
\label{fig:rp_mag_scatter}
\end{figure}

We compare this simple model for proximity zone scatter to the RT simulations in Figure~\ref{fig:rp_mag_scatter}. The median $R_p^{\rm alt}$, shown as the black solid curve and shaded region, almost exactly reproduces the dependence of $R_p^{\rm alt}$ with $M_{1450}$ from the RT simulations, shown in red. For very luminous quasars, the cumulative probability for $\tau_{\rm eff} > \tau_{\rm lim}$ in regions where $\bar{\tau}_{\rm eff}<\tau_{\rm lim}$ is enough to shift the median $R_p$ to lower values. Perhaps more interestingly, the scatter in $R_p^{\rm alt}$ is also well-reproduced by equation~(\ref{eqn:rpflucts2}), suggesting that the variation in $R_p^{\rm alt}$ between sightlines (and thus likely $R_p$ as well) is indeed dominated by the IGM scatter on the smoothing scale.

%%%%%%%%%%%%%%%%%%%%%%%%%%%%%%%%%%%%%%%%%%%%%%%%%%

% Don't change these lines
\bsp	% typesetting comment
\label{lastpage}
\end{document}